\documentclass[aps,prb,10pt,twocolumn,showpacs,showkeys,footinbib,superscriptaddress]{revtex4-1}

\usepackage[utf8x]{inputenc}
\usepackage{amsmath}
\usepackage{amssymb}
\usepackage{graphicx}
\usepackage[caption=false]{subfig}
\usepackage{pstricks}
\usepackage{color}
\usepackage{relsize}
\usepackage{bm}
\usepackage{braket}
\usepackage{epstopdf}

\usepackage{hyperref}
\usepackage[all]{hypcap}

\renewcommand{\i}{\ensuremath{\mathrm{i}}}
\newcommand{\e}{\ensuremath{\mathrm{e}}}
\renewcommand{\d}{\ensuremath{\mathrm{d}}}

\newcommand{\sgn}{\operatorname{{\mathrm sgn}}}

\newcommand{\B}{\mathbf}

\begin{document}
\title{Tuning Topological Superconductivity in Phase-Controlled Josephson Junctions with Rashba and Dresselhaus Spin-Orbit Coupling}

\author{Benedikt Scharf}
\affiliation{Institute for Theoretical Physics and Astrophysics and W\"{u}rzburg-Dresden Cluster of Excellence ct.qmat, University of W\"{u}rzburg, Am Hubland, 97074 W\"{u}rzburg, Germany}
\author{Falko Pientka}
\affiliation{Max Planck Institute for the Physics of Complex Systems, N\"{o}thnitzer Str. 38, 01187 Dresden, Germany}
\author{Hechen Ren}
\affiliation{Institute for Quantum Information and Matter, California Institute of Technology, Pasadena, California 91125, USA}
\author{Amir Yacoby}
\affiliation{Department of Physics, Harvard University, Cambridge, Massachusetts 02138, USA}
\author{Ewelina M. Hankiewicz}
\affiliation{Institute for Theoretical Physics and Astrophysics and W\"{u}rzburg-Dresden Cluster of Excellence ct.qmat, University of W\"{u}rzburg, Am Hubland, 97074 W\"{u}rzburg, Germany}

\date{\today}

\begin{abstract}
Recently, topological superconductors based on Josephson junctions in two-dimensional electron gases with strong Rashba spin-orbit coupling have been proposed as attractive alternatives to wire-based setups. Here, we elucidate how phase-controlled Josephson junctions based on quantum wells with [001] growth direction and an arbitrary combination of Rashba and Dresselhaus spin-orbit coupling can also host Majorana bound states for a wide range of parameters as long as the magnetic field is oriented appropriately. Hence, Majorana bound states based on Josephson junctions can appear in a wide class of two-dimensional electron gases. We study the effect of spin-orbit coupling, the Zeeman energies, and the superconducting phase difference to create a full topological phase diagram and find the optimal stability region to observe Majorana bound states in narrow junctions. Surprisingly, for equal Rashba and Dresselhaus spin-orbit coupling, well localized Majorana bound states can appear only for phase differences $\phi\neq\pi$ as the topological gap protecting the Majorana bound states vanishes at $\phi=\pi$. Our results show that the ratio between Rashba and Dresselhaus spin-orbit coupling or the choice of the in-plane crystallographic axis along which the superconducting phase bias is applied offer additional tunable knobs to test Majorana bound states in these systems. Finally, we discuss signatures of Majorana bound states that could be probed experimentally by tunneling conductance measurements at the edge of the junction.
\end{abstract}


\maketitle

\section{Introduction}\label{Sec:Intro}

The prospect of non-Abelian statistics and fault-tolerant quantum computing\cite{Nayak2008:RMP,Kitaev2003:AP,Alicea2011:NP} has made the search for Majorana bound states one of the most intense topics of research in condensed matter physics during the past decade.\cite{Kitaev2001:PhysUs,Alicea2012:RPP,Leijnse2012:SST,Beenakker2013:ARCMP} To realize such modes, it has long been proposed to use the midgap states of $p$-wave superconductors, where zero-energy Majorana bound states emerge at the edges or at vortices.\cite{Kopnin1991:PRB,Volovik1999:JETP,Senthil2000:PRB,Read2000:PRB,Kitaev2001:PhysUs,Sengupta2001:PRB} Although some superconducting materials, notably Sr$_2$RuO$_4$, are proposed to exhibit native $p$-wave symmetry,\cite{Mackenzie2003:RMP} topological superconductors, that is, superconductors exhibiting Majorana modes, can also be engineered from less exotic ingredients: They can emerge in materials with proximity-induced $s$-wave pairing and a nontrivial spin structure, typically provided by spin-orbit coupling (SOC) or magnetic textures.

\begin{figure}[t]
\centering
\includegraphics*[width=8.5cm]{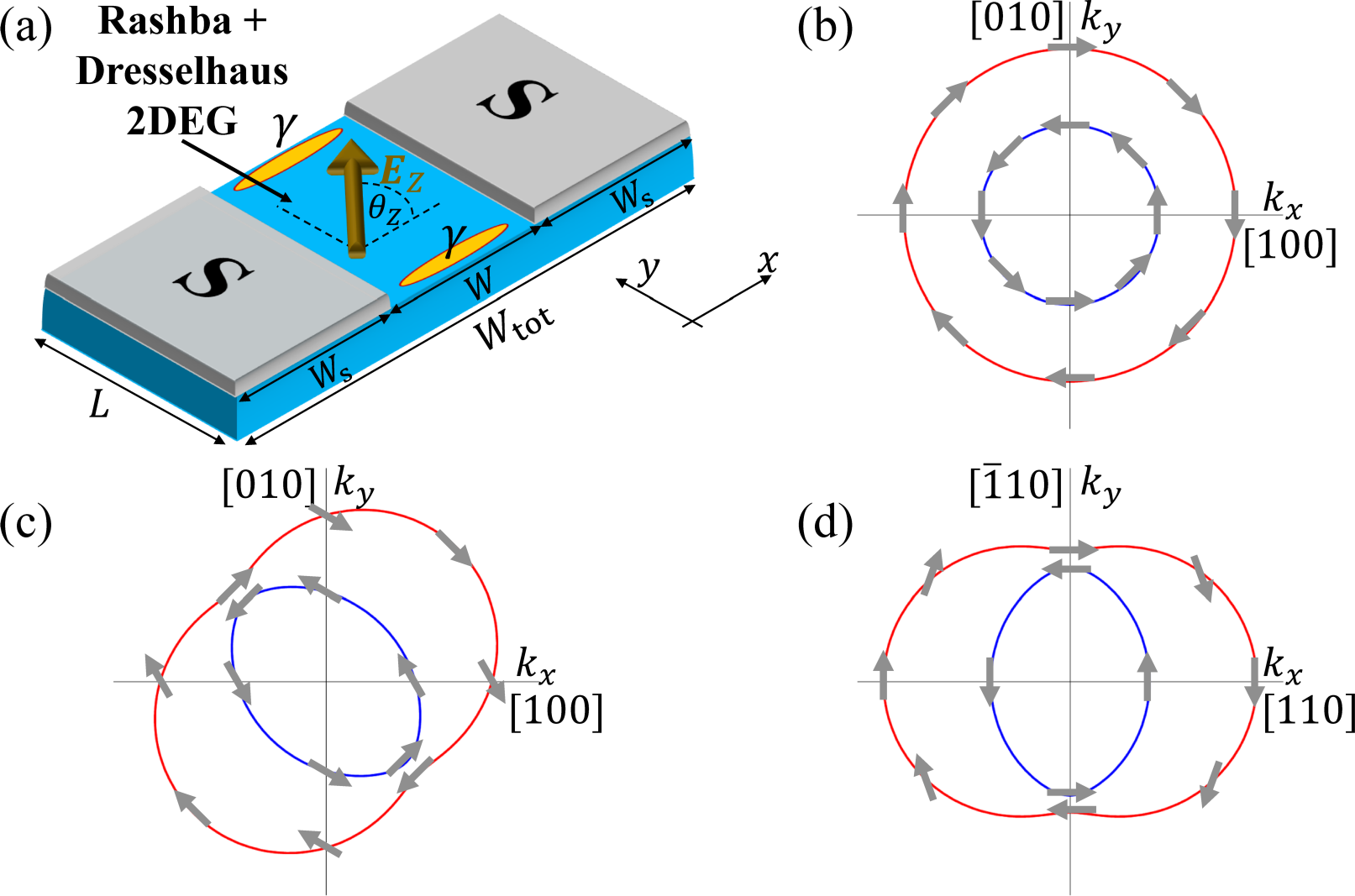}
\caption{(Color online) (a) Schematic setup of the phase-controlled Josephson junction. The position of the Majorana bound states $\gamma$ that appear at the ends of the normal region in the topological phase is also indicated. The angle of the direction of the in-plane Zeeman term with respect to the $x$ direction is given by $\theta_\mathrm{Z}$. Depending on the in-plane crystallographic axes along which the superconducting phase difference is applied, the SOC affects the formation of a topological phase differently. Fermi contours for 2DEGs formed in quantum wells with [001] growth direction: (b) 2DEG with Rashba SOC and 2DEGs with Rashba as well as Dresselhaus SOC if the $x$ axes are chosen along the (c) [100] and (d) [110] directions, respectively.}\label{fig:Scheme}
\end{figure}

In this context, experimental efforts have been mostly directed to one-dimensional (1D) systems such as hybrid structures of semiconductor nanowires and superconductors,\cite{Mourik2012:S,Das2012:NP,Deng2012:NL,Deng2016:S} where the interplay between proximity-induced $s$-wave superconductivity, a magnetic field and Rashba SOC results in a 1D topological superconductor.\cite{Lutchyn2010:PRL,Oreg2010:PRL,Pientka2012:PRL,Klinovaja2012:PRL,Kiczek2017:JPCM,Dominguez2017:NPJQM,Fleckenstein2018:PRB}\footnote{The use of magnetic textures instead of a constant magnetic field yields a synthetic SOC and removes the need for large Rashba SOC [M. Kjaergaard, K. W{\"o}lms, and K. Flensberg, Phys. Rev. B \textbf{85}, 020503(R) (2012); N. Mohanta, T. Zhou, J. Xu, J. E. Han, A. D. Kent, J. Shabani, I. \v{Z}uti\'{c}, A. Matos-Abiague, arXiv:1903.07834].} Despite promising experimental results, these 1D systems suffer from several drawbacks: Large Zeeman terms, that is, large $g$-factors and/or magnetic fields, and a good control of the chemical potential are required to drive the wire into its topological superconducting phase. Moreover, in order to harness Majorana bound states for topological quantum computing and to unambiguously prove their non-Abelian exchange statistics, braiding operations are required, that is, operations where different Majorana bounds states are exchanged with each other. As braiding statistics are ill-defined in 1D, an implementation of braiding operations necessitates complex wire networks instead of a single wire.\cite{Alicea2011:NP,Kim2015:PRB,Sau2011:PRB,Clarke2011:PRB,Halperin2012:PRB,Klinovaja2013:PRX}.

Recent experimental progress in proximity-inducing superconductivity in two-dimensional systems\cite{Hart2014:NP,Wan2015:NC,Shabani2016:PRB,Kjaergaard2016:NC,Hart2017:NP} or surface states,\cite{Maier2012:PRL,Sochnikov2015:PRL} however, points to a possible route for overcoming these obstacles, and different setups based on two-dimensional electron gases (2DEGs) have been proposed as alternatives to 1D wires.\cite{Sau2010:PRL,Alicea2010:PRB,Pientka2017:PRX,Hell2017:PRL,Virtanen2018:PRB,Fatin2016:PRL,*MatosAbiague2017:SSC,Zhou2019:PRB} Among these proposals, those based on phase-controlled Josephson junctions with Rashba SOC [see Fig.~\ref{fig:Scheme}(a)] offer an attractive alternative.\cite{Pientka2017:PRX,Hell2017:PRL,Ren2019:N,Fornieri2019:N,Stern2019:PRL,Setiawan2019:arxiv,Laeven2019:arxiv,Liu2019:PRB,Haim2019:PRL,Setiawan2019:PRB,Melo2019:arxiv} Here, the interplay between an in-plane Zeeman field parallel to the superconductor/normal (S/N) interfaces, Rashba SOC, and the Andreev bound states formed in the normal region induces topological superconductivity with Majorana bound states at the ends of the junction [see Fig.~\ref{fig:Scheme}(a)].\footnote{In the limit of narrow junctions, these 2DEGs can be viewed as a quasi-1D system along the S/N interfaces [that is, along the $y$ direction in Fig.~\ref{fig:Scheme}(a)]. However, the phase difference between the two S leads attached to the N region introduces a new ingredient that relaxes the constraints on the magnetic field and the chemical potential for the appearance of a topological phase.} A key advantage of this proposal is the tunable superconducting phase difference, which serves as an additional knob to control the topological transition. Moreover, the topological phase exists for a wider range of parameter values (chemical potential, magnetic field strength) than their wire-based counterparts. Hence, experimentally difficult fine-tuning like the one required in wires might not be needed.

In this paper, we show how the addition of Dresselhaus SOC in 2DEGs grown along the [001] crystallographic direction\footnote{By 2DEGs grown along the [001] crystallographic direction we mean 2DEGs where the [001] crystallographic direction is perpendicular to the plane of the 2DEG.} offers additional knobs that can be used to test and tune Majorana bound states in phase-controlled Josephson junctions: First, we show that Josephson junctions with an arbitrary combination of Rashba and Dresselhaus SOC can also host Majorana bound states as long as the in-plane magnetic field is oriented appropriately. The combination of Rashba and Dresselhaus SOC can yield various Fermi contours and spin textures\cite{Zutic2004:RMP,Fabian2007:APS} [see Figs.~\ref{fig:Scheme}(b)-(d) for examples] and introduces an effective spin-orbit field for propagation parallel to the S/N interfaces. Analogously to the situation in wires with SOC, the in-plane magnetic field should ideally be perpendicular to this effective spin-orbit field for a topological phase with localized Majorana end states to appear. Hence, Majorana bound states based on phase-controlled Josephson junctions can appear in a wide class of 2DEGs as long as there is strong SOC.\footnote{The interplay between Rashba and Dresselhaus SOC offers rich possibilities for topological states in superconducting systems: In 2D noncentrosymmetric superconductors, for example, the combination of Rashba and Dresselhaus SOC can give rise to topological edge states [M. Biderang, H. Yavari, M.-H. Zare, P. Thalmeier, and A. Akbari, Phys. Rev. B \textbf{98}, 014524 (2018)]. In proximity-induced superconducting [110] thin films with Dresselhaus SOC or in proximity-induced superconducting [001] thin films with equal Rashba and Dresselhaus SOC, flat zero-energy states emerge under an in-plane Zeeman field. [S. Ikegaya, Y. Asano, and Y. Tanaka, Phys. Rev. B \textbf{91}, 174511 (2015); S. Ikegaya and Y. Asano, Phys. Rev. B \textbf{95}, 214503 (2017); S. Ikegaya, S. Kobayashi, and Y. Asano, Phys. Rev. B \textbf{97}, 174501 (2018)].}

Second, we investigate the stability of the topological phase with respect to the strength and nature of SOC, the magnetic field direction and the superconducting phase difference $\phi$. Although the boundaries of the topological phase in narrow junctions do not depend strongly on SOC as long as the in-plane Zeeman field is oriented appropriately, the size of the topological gap protecting the Majorana bound states at a given $\phi$ sensitively depends on the exact combination of Rashba and Dresselhaus SOC. Surprisingly, for equal Rashba and Dresselhaus SOC, well-localized Majorana bound states can appear only for phase differences $\phi\neq\pi$ as the topological gap vanishes at $\phi=\pi$. We elucidate the origin of such gap closings and predict parameter ranges where they can be avoided and well-localized Majorana end states appear.

The addition of Dresselhaus SOC makes the system also sensitive to the choice of the in-plane axis along which the supercurrent flows. By keeping a [001] quantum-well growth direction, but rotating the junction setup in-plane, one can affect the formation of Majorana bound states [compare Figs.~\ref{fig:Scheme}(c) and (d)]: The conditions for the appearance of robust Majorana bound states differ for Josephson junctions with phase bias along the [100] direction and those with phase bias along the [110] direction. Our results imply that a tunable ratio between Rashba and Dresselhaus SOC offers an additional knob to probe the appearance and disappearance of Majorana bound states.

The manuscript is organized as follows: Section~\ref{Sec:Model} introduces the effective model used to describe the Josephson junctions with phase bias along the crystallographic [100] direction. This model is then used for an infinitely long Josephson junction to elucidate the conditions for a topological phase in Sec.~\ref{Sec:TP100}. In Sec.~\ref{Sec:TP110}, we then extend this discussion to how conditions for a topological phase are modified in Josephson junctions with phase bias along the [110] direction. Having determined the topological phase diagrams in such a way, we then turn to confined systems to explicitly demonstrate the appearance of Majorana bound states in Sec.~\ref{Sec:MBS}. Experimentally accessible signatures of these states and the optimal conditions under which Majorana bound states can be observed are discussed in Sec.~\ref{Sec:ExpSig}. Finally, we discuss schemes to tune and test topological superconductivity by Rashba and Dresselhaus SOC in Sec.~\ref{Sec:Tuning}. A brief summary in Sec.~\ref{Sec:Conclusions} concludes the paper. Readers that are mainly interested in experimentally observable signatures or schemes to use the interplay between Rashba and Dresselhaus SOC for tuning Majorana bound states may jump directly to Secs.~\ref{Sec:ExpSig} and~\ref{Sec:Tuning}.

\section{Model}\label{Sec:Model}

We consider a Josephson junction\footnote{Setups based on Josephson junctions, such as double-quantum-dot Josephson junctions, also offer intriguing prospects for Cooper pair splitters [R. S. Deacon, A. Oiwa, J. Sailer, S. Baba, Y. Kanai, K. Shibata, K. Hirakawa, and S. Tarucha, Nat. Commun. \textbf{6}, 7446 (2015); B. Probst, F. Dominguez, A. Schroer, A.L. Yeyati, and P. Recher, Phys. Rev. B \textbf{94}, 155445 (2016)].} based on a 2DEG with strong SOC, situated in the $xy$ plane and subject to an in-plane magnetic field,\footnote{Magnetic proximity effects [I. \v{Z}uti\'{c}, A. Matos-Abiague, B. Scharf, H. Dery, and K. Belashchenko, Materials Today \textbf{22}, 85 (2019)] offer a possible alternative to applying magnetic fields to induce an in-plane Zeeman field.} as depicted in Fig.~\ref{fig:Scheme}(a). The heterostructure in which the 2DEG forms is assumed to be grown in the crystallographic [001] direction. In our setup, the direction of the superconducting phase difference is denoted as the $x$ direction. For now, we assign the $x$ direction to the crystallographic [100] direction, but we will also discuss phase bias along other crystallographic directions in Sec.~\ref{Sec:TP110}. The pairing in the 2D S regions is induced from a nearby $s$-wave superconductor. With the basis order $\left(\hat{\psi}_\uparrow,\hat{\psi}_\downarrow,\hat{\psi}^\dagger_\downarrow,-\hat{\psi}^\dagger_\uparrow\right)$, the Bogoliubov-de Gennes (BdG) Hamiltonian describing this phase-controlled Josephson junction is then given by
\begin{equation}\label{eq:BDGHam}
\begin{array}{l}
\hat{H}_\mathrm{BdG}=\left[\frac{\hat{p}_x^2+\hat{p}_y^2}{2m}+\frac{\alpha}{\hbar}\left(s_y\hat{p}_x-s_x\hat{p}_y\right)+\frac{\beta}{\hbar}\left(s_x\hat{p}_x-s_y\hat{p}_y\right)\right.\\
\quad\quad\quad\quad\left.+\frac{m\alpha^2}{2\hbar^2}+\frac{m\beta^2}{2\hbar^2}-\mu_\mathrm{S}\right]\tau_z+\left(V_0\tau_z-\bm{E}_\mathrm{Z}\cdot\bm{s}\right)h(x)\\
\quad\quad\quad\quad+\Delta(x)\left[\tau_x\cos\Phi(x)-\tau_y\sin\Phi(x)\right].
\end{array}
\end{equation}
We will consider two different scenarios for a junction: (a) a junction with a finite width $W$ described by $h(x)=\Theta(W/2-|x|)$ and $\Delta(x)=\Delta\Theta(|x|-W/2)$, and (b) a $\delta$-barrier junction with $h(x)=W\delta(x)$ and $\Delta(x)=\Delta$. To describe the superconducting phase difference $\phi$ between the two S regions, we use the phase convention $\Phi(x)=(\pi-\phi)/2+\Theta(x)\phi$ for the finite and $\delta$-barriers. In Eq.~(\ref{eq:BDGHam}), $s_i$ and $\tau_i$ (with $i=x,y,z$) denote Pauli matrices in spin and particle-hole space, respectively. For brevity, we have not explicitly written the corresponding unit matrices $s_0$ and $\tau_0$ in Eq.~(\ref{eq:BDGHam}). Moreover, $\hat{p}_i$ denotes the momentum operator ($i=x,y$), $m$ the effective mass of conduction band electrons, $\Delta$ the induced pairing amplitude in the S regions, and $\bm{E}_\mathrm{Z}=(E_{\mathrm{Z},x},E_{\mathrm{Z},y},E_{\mathrm{Z},z})$ and $V_0$ are the Zeeman term and potential in the N region. Note that the potential $V_0$ can also be viewed as describing the difference between the chemical potentials in the S and N regions, $\mu_\mathrm{S}$ and $\mu_\mathrm{N}=\mu_\mathrm{S}-V_0$.

The strength and sign of Rashba and Dresselhaus SOC are given by $\alpha$ and $\beta$, respectively. In our description of Dresselhaus SOC, we omit contributions from the cubic Dresselhaus term, which is typically small compared to the linear term, especially if one of the momentum components, either $\hat{p}_x$ or $\hat{p}_y$, is small.\cite{Fabian2007:APS} As will be explained below in Sec.~\ref{Sec:TP100}, the topological phase boundaries are determined by setting $\hat{p}_y$ to zero. Conversely, the size of the gap protecting the topological superconducting phase is determined at momenta for which $\hat{p}_x$ is typically small. Therefore, we do not expect cubic Dresselhaus terms to play an important role and concentrate only on the linear Dresselhaus term. For an additional discussion, we refer to the end of Sec.~\ref{Sec:Finiteky}.

Regarding the dimensions of the Josephson junction, we consider S regions which are either semi-infinite or of finite width $W_\mathrm{S}$ in the $x$ direction. Typically, we determine the topological properties, such as the topological gap or the phase diagram (see Secs.~\ref{Sec:TP100} and~\ref{Sec:TP110} below), from systems that are infinite in the $y$ direction. To explicitly obtain Majorana bound states, however, we have to confine the system in $y$ direction, thereby introducing a finite length $L$ in Sec.~\ref{Sec:MBS}.

In the following, we will use either a scattering approach (see Appendix~\ref{Sec:Scatt}) or a finite-difference scheme (see Appendix~\ref{Sec:FD}) to solve the BdG equation
\begin{equation}\label{eq:BDG}
\hat{H}_\mathrm{BdG}\Psi(x,y)=E\Psi(x,y),
\end{equation}
where $\hat{H}_\mathrm{BdG}$ is given by Eq.~(\ref{eq:BDGHam}). In this way, we obtain the eigenstates $\Psi(x,y)$ and their corresponding eigenenergies $E$. While the scattering approach enables us to obtain analytical results in certain limiting cases and to gain some additional insight, the problem in general has to be solved numerically. Hence, if not explicitly stated otherwise, we employ the finite-difference method to solve Eq.~(\ref{eq:BDG}) with hard-wall boundary conditions at the ends of the structure at $x=\pm(W/2+W_\mathrm{S})$.

These finite-difference calculations are performed for a Josephson junction with finite widths of the S regions of $W_\mathrm{S}=450$ nm and a N region of finite width $W=100$ nm. Only when discussing differences between narrow and wide junctions in Sec.~\ref{Sec:ExpSig}, results for larger values of $W$ are also shown. Throughout the manuscript, the other parameters are chosen as $m=0.038m_0$ with the free electron mass $m_0$, $\mu_\mathrm{S}=1$ meV, $\mu_\mathrm{N}=0.7$ meV, and $\Delta=250$ $\mu$eV. Moreover, the total strength of SOC is fixed at $\lambda_\mathrm{soc}=16$ meVnm. Although we chose parameters characteristic of HgTe quantum wells (effective masses $m$, total strength of SOC $\lambda_\mathrm{soc}$), our conclusions are valid for many materials with strong SOC, such as InAs or InSb quantum wells, as shown in Sec.~\ref{Sec:Tuning}.\footnote{Here, we have chosen relatively small values for the chemical potentials $\mu_\mathrm{S}$ and $\mu_\mathrm{N}$ to allow for larger finite step sizes in our finite-difference method. However, going to larger chemical potentials does not qualitatively alter the results presented in this manuscript. Larger chemical potentials or SOC strengths typically lead to additional local extrema (or gap closings) of the Andreev spectrum compared to the spectra shown in Fig.~\ref{fig:Finiteky}. Depending on the material and the nearby superconductor used to proximity-induce superconductivity, induced superconducting gaps between $\Delta\approx60$ $\mu$eV and $\Delta\approx250$ $\mu$eV have been reported [see Refs.~\onlinecite{Mourik2012:S,Ren2019:N,Fornieri2019:N}]. Here, we use the upper value of $\Delta=250$ $\mu$eV, which leads to a faster exponential decay of the Andreev bound states in the S region.} Importantly, $|\beta|$ in InAs and InSb quantum wells is expected to be much larger compared to HgTe, where SOC is predominantly of Rashba-type, $|\beta|\ll|\alpha|$.

\section{Topological phase in Josephson junctions phase-biased along the [100] direction}\label{Sec:TP100}
Here, we briefly review the procedure to determine the appearance of a topological phase hosting Majorana bound states. As has been demonstrated in Ref.~\onlinecite{Pientka2017:PRX}, for $\alpha\neq0$ and $\beta=0$, an in-plane Zeeman term $E_{\mathrm{Z},y}$ parallel to the S/N interfaces (that is, in $y$ direction) puts the system in symmetry class BDI and can induce topological superconductivity. This topological phase in turn hosts Majorana bound states at the ends of the quasi-1D system (along the $y$ direction) formed by the N region, as illustrated in Fig.~\ref{fig:Scheme}(a). These Majorana bound states can be considered as being quasi-1D as long as the width $W$ of the normal region is smaller than the induced coherence length.\cite{Potter2010:PRL}\footnote{When $W$ becomes longer than the induced coherence length, the gap is exponentially suppressed. In that case, there can be multiple edge states at low energies which form a 1D propagating edge mode. In the system considered here, $W=100$ nm is much smaller than the coherence length $\xi\sim1$ $\mu$m and a quasi-1D description is thus appropriate.} The origin of the topological phase can be understood by considering an infinite Josephson junction in $y$ direction: In this case, $[\hat{H}_\mathrm{BdG},\hat{p}_y]=0$ and the momentum $k_y$ along the $y$ direction is a good quantum number. Then, the eigenstates in Eq.~(\ref{eq:BDG}) can be written as $\Psi(x,y)=\e^{\i k_yy}\psi_{k_y}(x)/\sqrt{S}$, where $S$ is the 2D unit area and $\psi_{k_y}(y)$ a 4-component spinor in Nambu space that is determined from $\hat{H}_\mathrm{BdG}(k_y)\psi_{k_y}(x)=E\psi_{k_y}(x)$. Here, $\hat{H}_\mathrm{BdG}(k_y)$ is given by Eq.~(\ref{eq:BDGHam}) with the operator $\hat{p}_y$ replaced by $\hbar k_y$. From the eigenspectrum $E(k_y=0,\phi)$ of $\hat{H}_\mathrm{BdG}(k_y=0)$ the ground-state parity can be determined as a function of the phase difference $\phi$ between the two S regions.\footnote{The quasi-1D system along the $y$ direction for arbitrary SOC and Zeeman field is in symmetry class D, for which the topological invariant is determined from the fermion parity of the ground state of $\hat{H}_\mathrm{BdG}(k_y=0)$, see Ref.~\onlinecite{Kitaev2001:PhysUs}.} In the absence of TRS-breaking, the spectrum is twofold degenerate and thus the ground-state parity is even. The Zeeman term $E_{\mathrm{Z},y}$ lifts this degeneracy and results in an odd ground-state parity for phases $\phi$ around $\phi=\pi$. It is this region around $\phi=\pi$ in $\phi$-space which supports a topological superconducting phase.\cite{Pientka2017:PRX} The values $\phi=\phi_c$ where the ground-state parity at $k_y=0$ changes are given by
\begin{equation}\label{eq:PhaseBoundary}
E(k_y=0,\phi=\phi_c)=0.
\end{equation}

Hence, the extent of the topological superconducting phase hosting Majorana bound states can be determined from (a) the zeros of $E(k_y=0,\phi)$, which determine the topological phase transitions, and (b) the existence of a gap $\Delta_\mathrm{top}$ in the eigenspectrum $E(k_y,\phi)$ inside the topological phase. Although the region defined by the phase boundaries is centered around $\phi=\pi$, $\Delta_\mathrm{top}$ is generally maximal for a superconducting phase difference $\phi\neq\pi$. A scattering matrix approach shows that only if the maximal value of $\Delta_\mathrm{top}$ approaches $\Delta$, one can typically expect this maximal $\Delta_\mathrm{top}$ to appear at $\phi=\pi$. In the following, we investigate the two conditions~(a) and~(b) for an arbitrary direction of the in-plane Zeeman field and an arbitrary combination of Rashba and Dresselhaus SOC to find a topological phase. As will be discussed below, the interplay between the Zeeman term and SOC is crucial for the opening of a topological gap.

\subsection{Topological phase boundaries}\label{Sec:ky0}
We first focus on topological phase transitions marked by gap closings at $k_y=0$. At $k_y=0$, the problem simplifies considerably because the Hamiltonian~(\ref{eq:BDGHam}) for an arbitrary combination of $\alpha$ and $\beta$ can be mapped to a Hamiltonian with $\beta=0$: First, we introduce the angles $\theta_\mathrm{soc}$ and $\theta_\mathrm{Z}$ to describe the combination of Rashba and Dresselhaus SOC and the direction of the in-plane Zeeman term via
\begin{equation}\label{eq:RDangle}
\begin{array}{l}
\alpha=\lambda_\mathrm{soc}\cos\theta_\mathrm{soc},\quad\beta=\lambda_\mathrm{soc}\sin\theta_\mathrm{soc}\\
\quad\quad\mathrm{with}\;\lambda_\mathrm{soc}=\sqrt{\alpha^2+\beta^2}
\end{array}
\end{equation}
and
\begin{equation}\label{eq:IPZeeman}
\begin{array}{l}
E_{\mathrm{Z},x}=E_\parallel\cos\theta_\mathrm{Z},\quad E_{\mathrm{Z},y}=E_\parallel\sin\theta_\mathrm{Z}\\
\quad\quad\mathrm{with}\;E_\parallel=\sqrt{E_{\mathrm{Z},x}^2+E_{\mathrm{Z},y}^2}.
\end{array}
\end{equation}
Here, we have parametrized the ratio between Rashba and Dresselhaus SOC by the angle $\theta_\mathrm{soc}$ and $\beta/\alpha=\tan\theta_\mathrm{soc}$ with $-\pi<\theta_\mathrm{soc}\leq\pi$. For general $\alpha$ and $\beta$, we can then perform the unitary transformation $\tilde{H}_\mathrm{BdG}(k_y=0)=\hat{U}(\theta_\mathrm{soc})\hat{H}_\mathrm{BdG}(k_y=0)\hat{U}^\dagger(\theta_\mathrm{soc})$, where $\hat{U}(\theta_\mathrm{soc})=\mathrm{diag}(1,\e^{\i\theta_\mathrm{soc}},1,\e^{\i\theta_\mathrm{soc}})$. Then, $\tilde{H}_\mathrm{BdG}(k_y=0)$ is given by Eq.~(\ref{eq:BDGHam}) with $\alpha\to\lambda_\mathrm{soc}$, $\beta\to0$, $E_{\mathrm{Z},x}\to E_\parallel\cos(\theta_\mathrm{Z}+\theta_\mathrm{soc})$ and $E_{\mathrm{Z},y}\to E_\parallel\sin(\theta_\mathrm{Z}+\theta_\mathrm{soc})$.

Hence, it is sufficient to investigate only the case of $\beta=0$ with arbitrary in-plane field directions when finding the zeros from Eq.~(\ref{eq:PhaseBoundary}). Any Zeeman term lifts the degeneracy between the two spin species and we typically find zero-energy crossings of $E(k_y=0,\phi)$ for all in-plane field directions. To corroborate this statement, we first consider a $\delta$-barrier junction with infinite superconducting leads, $W_\mathrm{S}\to\infty$. Then, the scattering approach allows us to obtain analytical results for the Andreev bound states $E(k_y=0,\phi)$, as explained in the Appendix~\ref{Sec:Scatt}. In the Andreev approximation, $\Delta\ll\mu_{S}$, we obtain the dispersion
\begin{equation}\label{eq:DAdisp}
\left|E(k_y=0,\phi)\right|\approx\frac{\Delta}{\sqrt{1+\left[f_\pm(Z_0,Z_\mathrm{Z},\phi)\right]^2}},
\end{equation}
where
\begin{equation}\label{eq:DAdef}
\begin{array}{l}
f_\pm(Z_0,Z_\mathrm{Z},\phi)=\\
\quad\quad\quad\quad\frac{2Z_\mathrm{Z}\pm\sqrt{2\left(1+Z_0^2+Z_\mathrm{Z}^2+(Z_\mathrm{Z}^2-Z_0^2)\cos\phi-\cos2\phi\right)}}{2+Z_0^2-Z_\mathrm{Z}^2+2\cos\phi}
\end{array}
\end{equation}
and $Z_\mathrm{Z}=\pi|\bm{E}_\mathrm{Z}|/E_\mathrm{T}$ and $Z_0=\pi V_0/E_\mathrm{T}$. Here, we have introduced the Thouless energy
\begin{equation}\label{eq:Thouless}
E_\mathrm{T}=\frac{\pi}{2}\frac{\hbar v_\mathrm{F}}{W}
\end{equation}
with the Fermi velocity $v_\mathrm{F}=\sqrt{2\mu_\mathrm{S}/m}$.

Equation~(\ref{eq:DAdisp}) contains several instructive features: First, it depends neither on SOC nor on the direction of the magnetic field. Second, its zeros, $\phi=\phi_c$, are given by the condition
\begin{equation}\label{eq:DAbc}
\cos\phi_c=-1+\frac{Z^2_\mathrm{Z}-Z^2_0}{2}.
\end{equation}
For $Z_\mathrm{Z}=0$ and finite $Z_0$, Eq.~(\ref{eq:DAbc}) cannot be satisfied and $E(k_y=0,\phi)$ exhibits no zero-energy crossing. Only if $|Z_\mathrm{Z}|>|Z_0|$ (or equivalently $|\bm{E}_\mathrm{Z}|>|V_0|$), (single) zeros are possible as the Zeeman term compensates for the normal reflection arising from the mismatch $V_0$ between the N and S regions. This competition between the Zeeman energy and the mismatch (or more generally normal reflection) yields a critical value of the Zeeman energy that needs to be exceeded for a topological transition to occur in non-transparent junctions. It is worth to mention here that for systems with finite $W_\mathrm{S}$ the finite size of the S regions can act as an additional source of normal reflection in the S/N/S junction. If $0<Z^2_\mathrm{Z}-Z^2_0\ll1$, the solution of Eq.~(\ref{eq:DAbc}) can be expanded and one obtains
\begin{equation}\label{eq:DAZ}
\phi_c\approx\pi\pm\pi\frac{\sqrt{|\bm{E}_\mathrm{Z}|^2-V_0^2}}{E_\mathrm{T}}+\mathcal{O}\left(\frac{\sqrt{|\bm{E}_\mathrm{Z}|^2-V_0^2}^3}{E_\mathrm{T}^3}\right).
\end{equation}
The two zero-energy crossings given by Eq.~(\ref{eq:DAZ}) define a range of $\phi$ values around $\phi=\pi$ with width $2\pi\sqrt{|\bm{E}_\mathrm{Z}|^2-V_0^2}/E_\mathrm{T}$, which can potentially host a topological phase (see below). Crucially, the conditions~(\ref{eq:DAbc}) and~(\ref{eq:DAZ}) are far less restrictive than the corresponding conditions in wire-based topological superconductors.

Equations~(\ref{eq:DAdisp}) and~(\ref{eq:DAbc}) have been obtained for a simplified model assuming a $\delta$-like N region and employing the Andreev approximation. Here, one directly matches the wave functions of the two S regions at the interface and incorporates the effect of the $\delta$-like N region via a boundary condition for the derivatives of the wave functions (see Appendix~\ref{Sec:Scatt}). Because of this, effects due to the phase acquired by electrons and holes propagating through a finite N region cannot be found in the $\delta$-barrier model. The most important of these effects of finite N regions is that there is not only a single transition from the trivial to the topological phase, as implied by Eq.~(\ref{eq:DAbc}), which yields a single phase boundary $\phi_c(|\bm{E}_\mathrm{Z}|,V_0)$. Instead, there is an alternation between trivial and topological phases with multiple phase transitions.\cite{Pientka2017:PRX} Such an alternation between trivial and topological phases as $|\bm{E}_\mathrm{Z}|$ increases is not captured by the $\delta$-barrier model, which provides a good approximation to a junction with finite-width N region only for very narrow junctions and $|\bm{E}_\mathrm{Z}|\ll E_\mathrm{T}$ (see Appendix~\ref{Sec:Scatt}).

If we go beyond these approximations and to finite N and S regions, we still find non-degenerate zeros for sufficiently large $|\bm{E}_\mathrm{Z}|$. These zeros and the corresponding phase boundaries remain largely independent of SOC, consistent with Refs.~\onlinecite{Chtchelkatchev2003:PRL,Beri2008:PRB}, where it has been found that the Andreev spectrum around zero energy is not affected by SOC in short junctions. Now, however, $E(k_y=0,\phi)$ also slightly depends on the direction of the magnetic field.\footnote{Only for special combinations of the magnetic field and SOC, the SOC can be gauged away in Eqs.~(\ref{eq:BDGHam}) and~(\ref{eq:BDGHamRedefined}) at $k_y=0$.} As can be discerned from Eq.~(\ref{eq:DAZ}), the Thouless energy $E_\mathrm{T}$ sets the energy scale for the distance between the two zero-energy solutions. This is also the case for Josephson junctions with finite N regions. Now $E_\mathrm{T}$ is defined, however, from the Fermi velocity of the N region, that is, $v_\mathrm{F}=\sqrt{2\mu_\mathrm{N}/m}$. In the remainder of this work, we provide results for junctions with finite N regions. Our numerical results presented below for these finite junctions indeed exhibit additional phase transitions. Still, Eq.~(\ref{eq:DAdisp}) provides a good qualitative description of the topological phase transition at $|\bm{E}_\mathrm{Z}|<E_\mathrm{T}$ in very narrow junctions. We have included the discussion of the $\delta$-barrier model here because it allows for an understanding of the topological phase transition on a basic level.

\begin{figure}[t]
\centering
\includegraphics*[width=8.5cm]{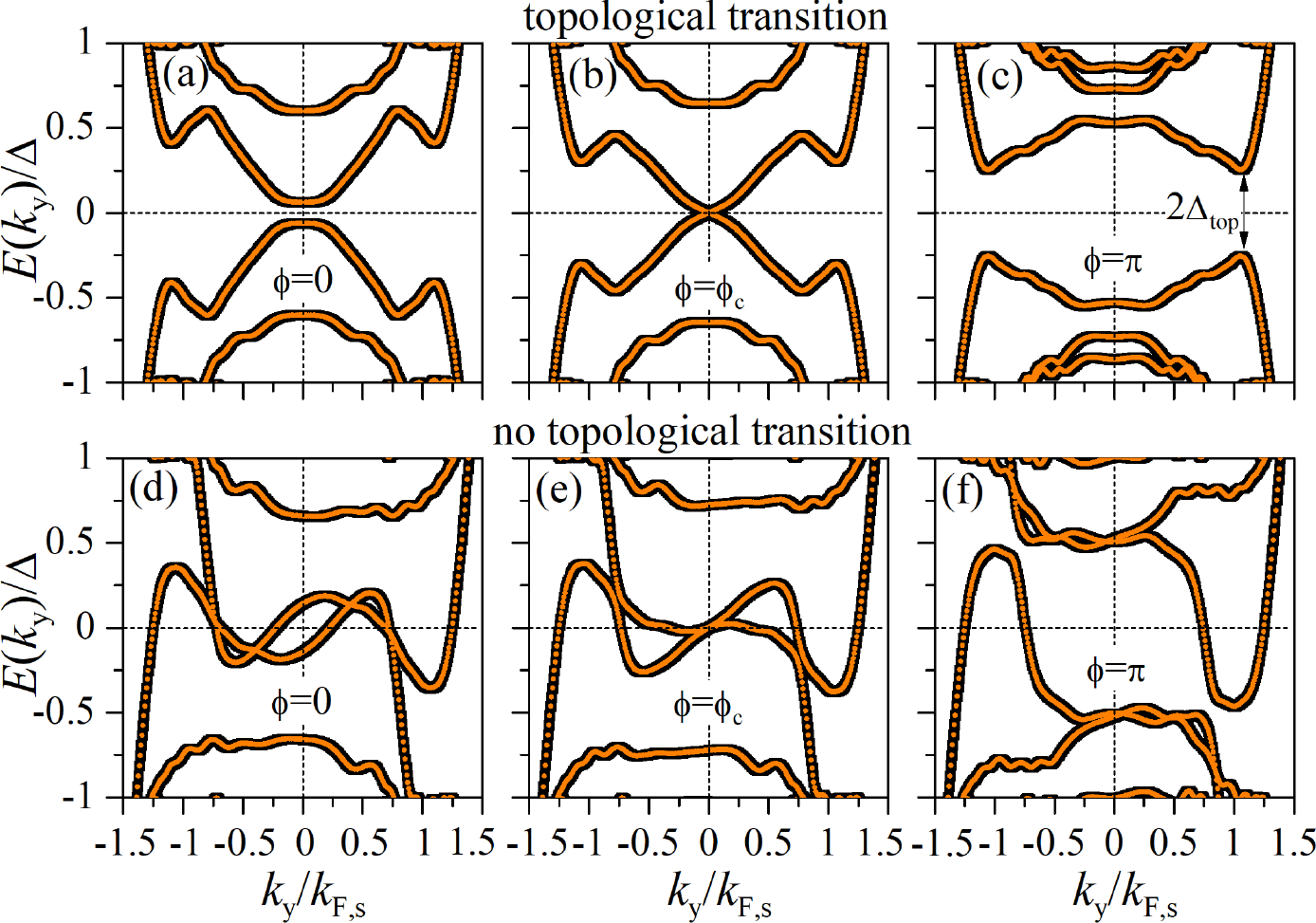}
\caption{(Color online) Andreev bound state spectrum as a function of the transverse momentum $k_y$ at (a,d) $\phi=0$, (b,e) $\phi=\phi_c$, given by Eq.~(\ref{eq:PhaseBoundary}), and (c,f) $\phi=\pi$. In panels~(a)-(c), either $\theta_\mathrm{soc}=0$ (only Rashba SOC, $\beta=0$), $\bm{E}_\mathrm{Z}=|\bm{E}_\mathrm{Z}|\bm{e}_y$ or $\theta_\mathrm{soc}=\pi/2$ (only Dresselhaus SOC, $\alpha=0$), $\bm{E}_\mathrm{Z}=|\bm{E}_\mathrm{Z}|\bm{e}_x$, that is, $\bm{E}_\mathrm{Z}\perp\bm{n}_\mathrm{soc}$. In panels~(d)-(f), either $\theta_\mathrm{soc}=0$, $\bm{E}_\mathrm{Z}=|\bm{E}_\mathrm{Z}|\bm{e}_x$ or $\theta_\mathrm{soc}=\pi/2$, $\bm{E}_\mathrm{Z}=|\bm{E}_\mathrm{Z}|\bm{e}_y$, that is, $\bm{E}_\mathrm{Z}\parallel\bm{n}_\mathrm{soc}$. The topological gap $\Delta_\mathrm{top}$ appearing at $\phi=\pi$ close to $k_\mathrm{F}+k_\mathrm{soc}$ is indicated in panel~(c). No finite topological gap $\Delta_\mathrm{top}$ arises in panel~(f), where for any $|E|<\Delta$ a state can be found. In all panels, $m=0.038m_0$, $W=100$ nm, $\lambda_\mathrm{soc}=\sqrt{\alpha^2+\beta^2}=16$ meVnm, $\mu_\mathrm{S}=1$ meV, $\mu_\mathrm{N}=0.7$ meV, $\Delta=250$ $\mu$eV, and $|\bm{E}_\mathrm{Z}|=0.5$ meV. The spectra have been computed employing a finite-difference method along the $x$ direction with a width of the entire S/N/S junction of $W_\mathrm{tot}=2W_\mathrm{S}+W=1$ $\mu$m.}\label{fig:Finiteky}
\end{figure}

\subsection{Subgap spectrum}\label{Sec:Finiteky}
In Sec.~\ref{Sec:ky0}, we have seen that non-degenerate zeros of $E(k_y=0,\phi)$ arise as $|\bm{E}_\mathrm{Z}|$ increases and overcomes the mismatch in chemical potentials between the N and S regions. This behavior is largely independent of the direction of the Zeeman field. Still, the appearance of a topological phase around $\phi=\pi$ that can host Majorana bound states also requires the spectrum at or around $\phi=\pi$ to be gapped for any $k_y$, not just for $k_y=0$.

Similar to wires,\cite{Lutchyn2010:PRL,Oreg2010:PRL} the interplay between the Zeeman term and SOC plays a crucial role in ensuring a gapped spectrum and thus a topological phase. In order to make that analogy to wires with strong SOC clearer and to understand under which conditions Majorana bound states appear, it is convenient to recast the Hamiltonian given by Eq.~(\ref{eq:BDGHam}): With Eq.~(\ref{eq:RDangle}), $k_\mathrm{soc}=m\lambda_\mathrm{soc}/\hbar^2$, the mean Fermi wave vectors $k_\mathrm{F,S}=\sqrt{2m\mu_\mathrm{S}}/\hbar$ and $k_\mathrm{F,N}=\sqrt{2m\mu_\mathrm{N}}/\hbar$ as well as the SO fields
\begin{equation}\label{eq:hnSOCo100}
\bm{h}_\mathrm{soc}=(\beta,\alpha,0)\quad\text{and}\quad\bm{n}_\mathrm{soc}=(\alpha,\beta,0),
\end{equation}
Eq.~(\ref{eq:BDGHam}) acquires the form
\begin{equation}\label{eq:BDGHamRedefined}
\begin{array}{l}
\hat{H}_\mathrm{BdG}=\\
\;\left[\frac{\hat{p}_x^2+\hat{p}_y^2+\hbar^2\left(k_\mathrm{soc}^2-k_\mathrm{F,S}^2\right)}{2m}+\left(\bm{h}_\mathrm{soc}\cdot\bm{s}\right)\frac{\hat{p}_x}{\hbar}-\left(\bm{n}_\mathrm{soc}\cdot\bm{s}\right)\frac{\hat{p}_y}{\hbar}\right]\tau_z\\
\;+\left(V_0\tau_z-\bm{E}_\mathrm{Z}\cdot\bm{s}\right)h(x)\\
\;+\Delta(x)\left[\tau_x\cos\Phi(x)-\tau_y\sin\Phi(x)\right],
\end{array}
\end{equation}
where $\bm{s}=(s_x,s_y,s_z)$ contains the Pauli spin matrices. As before, the operator $\hat{p}_y$ is replaced by $\hbar k_y$ for the infinite system.

In semiconductor-wire-based topological superconductors, a gapped spectrum arises if $\bm{E}_\mathrm{Z}$ is perpendicular to the SO field. If $\bm{E}_\mathrm{Z}$ is parallel to the SO field, on the other hand, the spectrum remains gapless and a finite wire cannot host Majorana bound states. In our quasi-1D system along the $y$ direction, the situation is very similar:\footnote{An alternative procedure for narrow Josephson junctions is to obtain an effective 1D Hamiltonian for the N region using a transfer matrix approach similar to the one presented in Ref.~\onlinecite{Virtanen2018:PRB}. In this approach, our setup can be considered as a wire with a Zeeman term and induced superconductivity.} For propagation parallel to the S/N interfaces (that is, along the $y$ direction), the relevant effective SO field is given by $\bm{n}_\mathrm{soc}$. Only for $\bm{E}_\mathrm{Z}\perp\bm{n}_\mathrm{soc}$ or $\bm{E}_\mathrm{Z}$ close to satisfying this condition, the spectrum $E(\phi\neq\phi_c,k_y)$ of the Josephson junction is gapped for any $k_y$.

This is illustrated in Fig.~\ref{fig:Finiteky}, which contains the energy spectra computed with the finite-difference method for a finite Josephson junction with fixed Zeeman energy $|\bm{E}_\mathrm{Z}|=0.5$ meV. Figures~\ref{fig:Finiteky}(a)-(c) show the topological phase transition occurring if $\bm{E}_\mathrm{Z}\perp\bm{n}_\mathrm{soc}$: The black lines show results if $\bm{E}_\mathrm{Z}=|\bm{E}_\mathrm{Z}|\bm{e}_y$ and only Rashba SOC is present in the system, $\theta_\mathrm{soc}=0$, while the orange lines show results for $\bm{E}_\mathrm{Z}=|\bm{E}_\mathrm{Z}|\bm{e}_x$ and $\theta_\mathrm{soc}=\pi/2$, that is, if there is only Dresselhaus SOC. In both cases, $E(k_y,\phi)$ is gapped for $\phi<\phi_c$ [Fig.~\ref{fig:Finiteky}(a)] and exhibits a gap closing at $k_y=0$ for $\phi=\phi_c$ [Fig.~\ref{fig:Finiteky}(b)]. For $\phi>\phi_c$, the gap is reopened as the Josephson junction enters the topological regime [Fig.~\ref{fig:Finiteky}(c)]. In Fig.~\ref{fig:Finiteky}(c), the topological gap
\begin{equation}\label{eq:TopGap}
\Delta_\mathrm{top}(\phi)=\underset{k_y}{\mathrm{min}}\left[|E(k_y,\phi)|\right],
\end{equation}
appearing close to $k_y\approx k_\mathrm{F,N}+k_\mathrm{soc}$ here, is also indicated for $\phi=\pi$.

If $\bm{E}_\mathrm{Z}\parallel\bm{n}_\mathrm{soc}$, the situation is very different as illustrated by Figs~\ref{fig:Finiteky}(d)-(f): We have chosen the same parameters as in Figs~\ref{fig:Finiteky}(a)-(c), but with $\theta_\mathrm{soc}=0$ and $\bm{E}_\mathrm{Z}=|\bm{E}_\mathrm{Z}|\bm{e}_x$ (black lines) or with $\theta_\mathrm{soc}=\pi/2$ and $\bm{E}_\mathrm{Z}=|\bm{E}_\mathrm{Z}|\bm{e}_y$ (orange lines). Although there is a phase $\phi=\phi_c$ where $E(k_y=0,\phi_c)$ vanishes, there are other zeros at finite $k_y$ and the spectrum is always gapless. Hence, no topological regime forms for $\phi>\phi_c$.

Qualitatively, the appearance of a gapped spectrum at finite $k_y$ can be understood in the following way: Extrema of $E(k_y,\phi)$ appear in the vicinity of $|k_y|\approx k_\mathrm{F,N}\pm k_\mathrm{soc}$. Close to these momenta $k_y$, the $\hat{p}_x$-terms in Eq.~(\ref{eq:BDGHamRedefined}) are much smaller than the $k_y$-dependent terms, and we will thus omit them for a moment. Then, only the interplay between $\bm{E}_\mathrm{Z}$ and $\bm{n}_\mathrm{soc}$ determines the opening of a gap around $E=0$. Similar to the case of proximitized nanowires, the magnetic field $\bm{E}_\mathrm{Z}$ needs to be perpendicular to the spin-orbit field, $\bm{E}_\mathrm{Z}\perp\bm{n}_\mathrm{soc}$, for a gap to open around $E=0$. As $\bm{E}_\mathrm{Z}$ is rotated away from $\bm{E}_\mathrm{Z}\perp\bm{n}_\mathrm{soc}$, the spectrum is no longer symmetric with respect to $k_y=0$, $E(k_y,\phi)\neq E(-k_y,\phi)$, but tilted [see, for example, Fig.~\ref{fig:Finiteky}, where the gap opening is moved to $E<0$ for $k_y>0$]. The effect is that $\Delta_\mathrm{top}$ quickly decreases and $\Delta_\mathrm{top}\to0$ as $\bm{E}_\mathrm{Z}$ is rotated away from $\bm{E}_\mathrm{Z}\perp\bm{n}_\mathrm{soc}$. In the above discussion, we have omitted $\bm{h}_\mathrm{soc}$ due to the small contribution arising from it at $|k_y|\approx k_\mathrm{F,N}\pm k_\mathrm{soc}$. This contribution can be viewed as an effective correction to the Zeeman term $\bm{E}_\mathrm{Z}$ and does not qualitatively change the argument presented above. Despite this analogy to proximitized nanowires, we emphasize again that the phase difference $\phi$ between the two S leads attached to the N region provides an additional knob that relaxes the constraints on $|\bm{E}_\mathrm{Z}|$, $\mu$, and $\Delta$ for a topological transition to occur.

Since the size of $\Delta_\mathrm{top}$ is typically determined at transverse momenta close to $|k_y|\approx k_\mathrm{F,N}\pm k_\mathrm{soc}$, where $k_x$ is small, we do not expect that corrections due to cubic Dresselhaus terms will significantly change the above behavior of $\Delta_\mathrm{top}$. Likewise, the topological phase boundaries $\phi_c$ are determined at $k_y=0$, where there is no contribution from cubic Dresselhaus terms. Hence, the phase boundaries are not affected by cubic Dresselhaus terms. This is the reason why we restrict ourselves to only linear Dresselhaus terms in this work.

\subsection{Topological phase diagram}\label{Sec:TopPD}
\begin{figure}[t]
\centering
\includegraphics*[width=8.5cm]{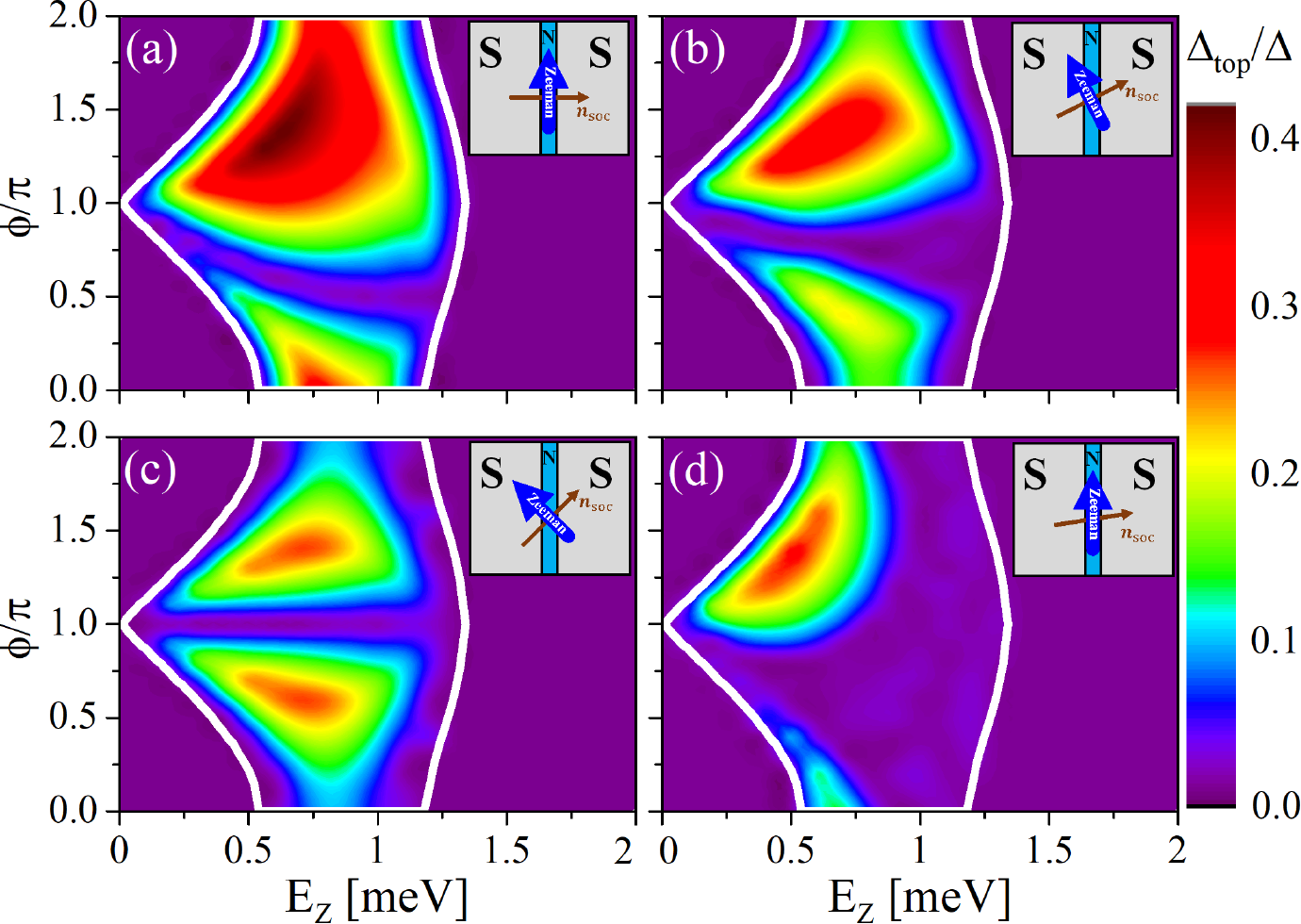}
\caption{(Color online) Topological phase diagrams for different configurations of SOC, described by $\theta_\mathrm{soc}$, and the Zeeman term $\bm{E}_\mathrm{Z}$: (a) $\theta_\mathrm{soc}=0$ ($\alpha=16$ meVnm, $\beta=0$), $\bm{E}_\mathrm{Z}=|\bm{E}_\mathrm{Z}|\bm{e}_y$, (b) $\theta_\mathrm{soc}=0.15\pi$ ($\alpha\approx14.3$ meVnm, $\beta\approx7.3$ meVnm), $\bm{E}_\mathrm{Z}\perp\bm{n}_\mathrm{soc}$, (c) $\theta_\mathrm{soc}=0.25\pi$ ($\alpha=\beta\approx11.3$ meVnm), $\bm{E}_\mathrm{Z}\perp\bm{n}_\mathrm{soc}$ and (d) $\theta_\mathrm{soc}=0.05\pi$ ($\alpha\approx15.8$ meVnm, $\beta\approx2.5$ meVnm), $\bm{E}_\mathrm{Z}=|\bm{E}_\mathrm{Z}|\bm{e}_y$. The total strength of SOC is $\lambda_\mathrm{soc}=\sqrt{\alpha^2+\beta^2}=16$ meVnm in all panels and all other parameters are the same as in Fig.~\ref{fig:Finiteky}. The white lines indicate the phase boundaries $\phi_c$ computed from Eq.~(\ref{eq:PhaseBoundary}) and $\Delta_\mathrm{top}$ only measures the gap inside the topological phase, whereas gaps outside the topological phase have been set to zero. The insets illustrate the directions of $\bm{n}_\mathrm{soc}$ and $\bm{E}_\mathrm{Z}$.}\label{fig:PhaseDiagram}
\end{figure}

From Secs.~\ref{Sec:ky0} and~\ref{Sec:Finiteky}, we conclude that a topological phase can arise for extended regions in $\bm{E}_\mathrm{Z}$-$\phi$-space as long as $\bm{E}_\mathrm{Z}\perp\bm{n}_\mathrm{soc}$ and $|\bm{E}_\mathrm{Z}|$ is large enough to overcome normal reflection. This can be seen in Fig.~\ref{fig:PhaseDiagram}, which shows the zero-energy crossing at $k_y=0$ from Eq.~(\ref{eq:PhaseBoundary}) (white lines) as well as the topological gap $\Delta_\mathrm{top}$ at finite $k_y$ from Eq.~(\ref{eq:TopGap}) (color map).\footnote{For $(\bm{E}_\mathrm{Z},\phi)$ outside the topological region, Eq.~(\ref{eq:TopGap}) can also yield a gap. This gap, however, is trivial and thus, we set $\Delta_\mathrm{top}=0$ outside the topological region.} The results are presented for the Josephson junction from Fig.~\ref{fig:Finiteky} for different configurations of SOC and $\bm{E}_\mathrm{Z}$.

Figure~\ref{fig:PhaseDiagram}(a) depicts the phase diagram discussed in Ref.~\onlinecite{Pientka2017:PRX} for the case of pure Rashba SOC with $\alpha\neq0$, $\beta=0$ and $\bm{E}_\mathrm{Z}=|\bm{E}_\mathrm{Z}|\bm{e}_y$. Here, large parts of the region defined by the boundaries $\phi_c$ from Eq.~(\ref{eq:PhaseBoundary}) exhibit a sizable gap of a few tenths of $\Delta$. The region defined by the boundaries $\phi_c$ is centered around $\phi=\pi$, and a finite gap $\Delta_\mathrm{top}$ exists at $\phi=\pi$. As mentioned before, however, this does not imply that $\Delta_\mathrm{top}$ is necessarily always maximal at $\phi=\pi$. In fact, the maximal value of $\Delta_\mathrm{top}$ in Fig.~\ref{fig:PhaseDiagram}(a) is reached for $\phi>\pi$. On the other hand, in a narrow range of phases $\phi<\pi$, gap closings appear inside the topological phase. The extension of the gapless phase slightly increases at larger Zeeman fields. While the phase diagram shows only one topological phase, additional phases exist at higher Zeeman energies in agreement with the discussion in Sec.~\ref{Sec:ky0}. The next topological phase transition occurs at $E_Z\simeq4$ meV (not shown).

If both Rashba and Dresselhaus SOC are present, an extended topological phase in $\bm{E}_\mathrm{Z}$-$\phi$-space can also be achieved as long as $\bm{E}_\mathrm{Z}\perp\bm{n}_\mathrm{soc}$. This is illustrated in Figs.~\ref{fig:PhaseDiagram}(b) and~(c), where $\theta_\mathrm{soc}=0.15\pi$ and $\theta_\mathrm{soc}=0.25\pi$, respectively, and $\bm{E}_\mathrm{Z}\perp\bm{n}_\mathrm{soc}$. Similar to Fig.~\ref{fig:PhaseDiagram}(a), nearly the entire region defined by the boundaries $\phi_c$ from Eq.~(\ref{eq:PhaseBoundary}) can host a topological phase with $\Delta_\mathrm{top}\neq0$, although the gap is somewhat reduced compared to the case of pure Rashba SOC. Interestingly, in the case $\alpha=\beta$ [see Fig.~\ref{fig:PhaseDiagram}(c)], the gapless regime is restricted to $\phi=\pi$ and $\Delta_\mathrm{top}$ is symmetric under $\phi\to\pi-\phi$. We elucidate the occurrence of these gap closings based on symmetry considerations in Sec.~\ref{Sec:TopGap}.

In contrast to Figs.~\ref{fig:PhaseDiagram}(a)-(c), the regions in $\bm{E}_\mathrm{Z}$-$\phi$-space with $\Delta_\mathrm{top}\neq0$ are greatly reduced if $\bm{E}_\mathrm{Z}\perp\bm{n}_\mathrm{soc}$ is not satisfied. Such a situation is illustrated by Fig.~\ref{fig:PhaseDiagram}(d), where $\bm{E}_\mathrm{Z}$ is slightly misaligned from $\bm{E}_\mathrm{Z}\perp\bm{n}_\mathrm{soc}$. Here, $\bm{E}_\mathrm{Z}$ is kept parallel to the S/N interfaces, $\bm{E}_\mathrm{Z}=|\bm{E}_\mathrm{Z}|\bm{e}_y$, but there is a combination of Dresselhaus and dominant Rashba SOC ($\theta_\mathrm{soc}=0.05\pi$). First of all, one can observe that the boundaries $\phi_c$ given by Eq.~(\ref{eq:PhaseBoundary}) are nearly the same as in Fig.~\ref{fig:PhaseDiagram}(a), also consistent with our findings from the $\delta$-barrier model for narrow junctions where $E(k_y=0,\phi)$ does not depend on SOC or on the direction of $\bm{E}_\mathrm{Z}$. The behavior of $\Delta_\mathrm{top}$, on the other hand, is very different compared to the case of pure Rashba SOC and a topological phase with a sizable $\Delta_\mathrm{top}$ arises only in a much reduced region close to the boundary $\phi_c<\pi$.

This reduction of the topological phase with finite $\Delta_\mathrm{top}$ is due to the gap closing mechanisms discussed in Sec.~\ref{Sec:Finiteky} if $\bm{E}_\mathrm{Z}\perp\bm{n}_\mathrm{soc}$ is not fulfilled. If $\bm{E}_\mathrm{Z}$ deviates by too much from $\bm{E}_\mathrm{Z}\perp\bm{n}_\mathrm{soc}$, $\Delta_\mathrm{top}\to0$ for all values in $\bm{E}_\mathrm{Z}$-$\phi$-space and hence no topological phase forms at all. This can in turn be used to define a critical angle describing by how much $\bm{E}_\mathrm{Z}$ and $\bm{n}_\mathrm{soc}$ may deviate from the condition $\bm{E}_\mathrm{Z}\perp\bm{n}_\mathrm{soc}$ before the topological phase disappears altogether. For the parameters and field direction in Figs.~\ref{fig:PhaseDiagram}(a) and~(d), the disappearance of the topological phase occurs for a deviation of around $12^\circ\simeq0.07\pi$ from $\bm{E}_\mathrm{Z}\perp\bm{n}_\mathrm{soc}$. Calculations with InAs or InSb parameters also show that the topological phase typically vanishes if $\bm{E}_\mathrm{Z}$ deviates from $\bm{E}_\mathrm{Z}\perp\bm{n}_\mathrm{soc}$ by more than around $10^\circ\simeq0.06\pi$.

\subsection{Size of $\Delta_\mathrm{top}$ and possible gap closings}\label{Sec:TopGap}

\subsubsection{General considerations}

The appearance and size of a topological gap $\Delta_\mathrm{top}$ is crucial for the localization and stability of a Majorana bound state arising in the S/N/S junction. In contrast to the boundaries $\phi_c$, $\Delta_\mathrm{top}$ can very much depend on the particular form of SOC even if $\bm{E}_\mathrm{Z}$ is kept perpendicular to $\bm{n}_\mathrm{soc}$ in narrow junctions with $\Delta\ll E_\mathrm{T}$. Hence, the size of $\Delta_\mathrm{top}$ is studied in more detail here.

One limiting factor, already noted in Ref.~\onlinecite{Pientka2017:PRX} and independent of SOC, is the width $W$ of the N region: As $W$ increases, the number of Andreev bound-state bands in the subgap spectrum ($|E|<\Delta$) increases, pushing the lowest Andreev band closer to $E=0$, thereby also reducing $\Delta_\mathrm{top}$. In the optimal case, $\Delta_\mathrm{top}$ can at most be of the order of $\hbar^2/mW^2$, provided $\hbar^2/mW^2\leq\Delta$. This implies that narrow junctions, such as the ones studied above with $W=100$ nm, are beneficial to observe well-localized Majorana bound states. Whereas the upper limit for $\Delta_\mathrm{top}$ is governed by $W$, the actual value of $\Delta_\mathrm{top}$ depends on other system parameters such as SOC, $\bm{E}_\mathrm{Z}$ or normal reflection.

While $\mu_\mathrm{N}$ can affect $\Delta_\mathrm{top}$ and determines the center $|\bm{E}_\mathrm{Z}|\approx E_\mathrm{T}=(\pi/2)\hbar\sqrt{2\mu_\mathrm{N}/mW^2}$ of the topological region, its effect is typically small if normal reflection is weak. The topological gap does, however, sensitively depend on the combination of Rashba and Dresselhaus SOC and the direction of $\bm{E}_\mathrm{Z}$.

\subsubsection{Dependence of $\Delta_\mathrm{top}$ on $\theta_\mathrm{soc}$}

\begin{figure}[t]
\centering
\includegraphics*[width=8.5cm]{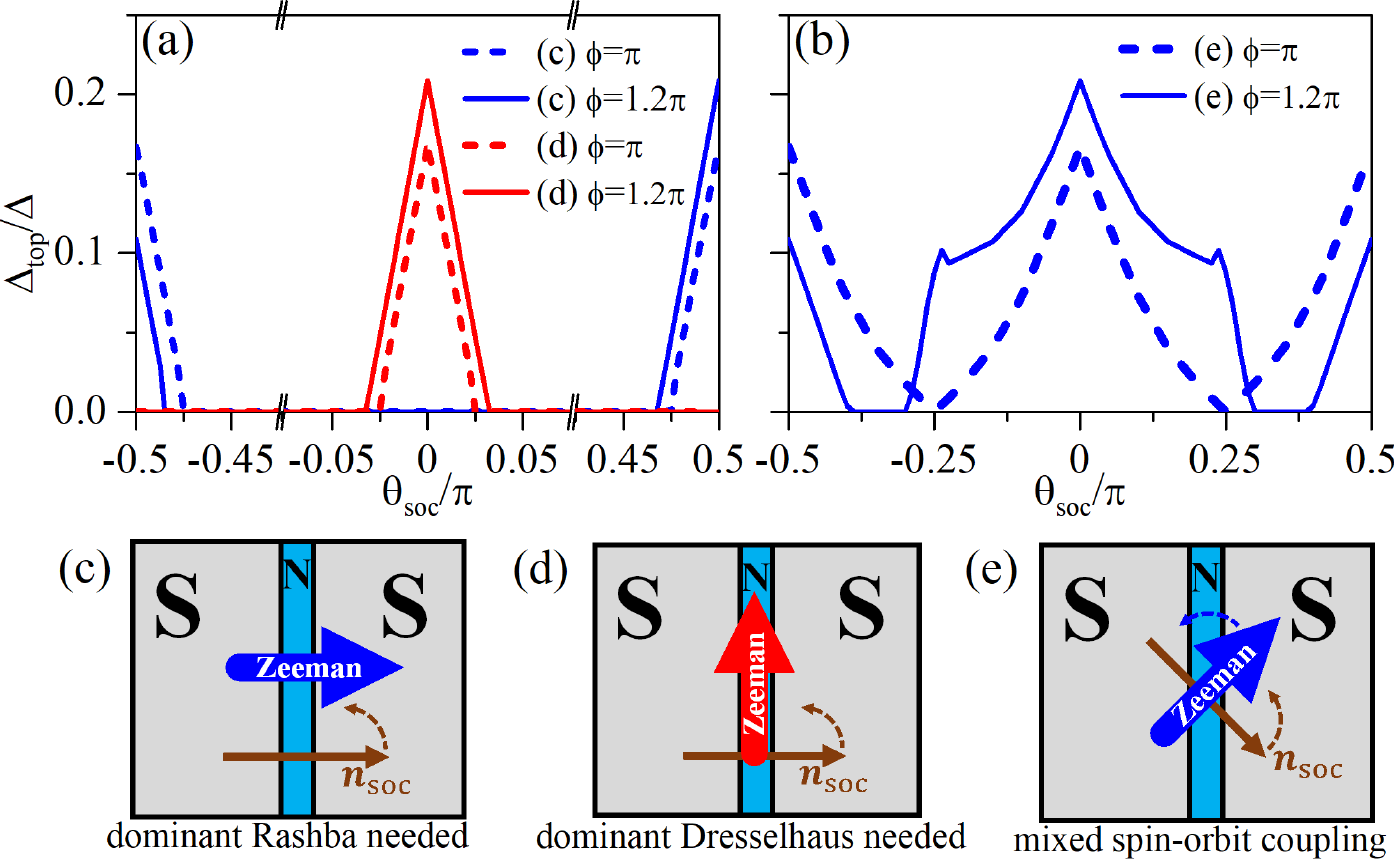}
\caption{(Color online) Dependence of the topological gap $\Delta_\mathrm{top}$ at $\phi=\pi$ (dashed curves) and $\phi=1.2\pi$ (solid curves) on $\theta_\mathrm{soc}$ for different configurations of $\bm{E}_\mathrm{Z}$: (a) $\bm{E}_\mathrm{Z}=|\bm{E}_\mathrm{Z}|\bm{e}_x$ (blue) and $\bm{E}_\mathrm{Z}=|\bm{E}_\mathrm{Z}|\bm{e}_y$ (red) as well as (b) $\bm{E}_\mathrm{Z}$ rotated with $\theta_\mathrm{soc}$ such that $\bm{E}_\mathrm{Z}\perp\bm{n}_\mathrm{soc}$ (blue). The parameters are the same as in Fig.~\ref{fig:Finiteky} with $|\bm{E}_\mathrm{Z}|=1.1$ meV kept constant for the different configurations of $\bm{E}_\mathrm{Z}$. The lower panels (c)-(e) show schemes of the three different configurations presented in panels~(a) and~(b).}\label{fig:TopGapky}
\end{figure}

The dependence of $\Delta_\mathrm{top}$ on $\theta_\mathrm{soc}$ for fixed $\lambda_\mathrm{soc}=16$ meVnm and $|\bm{E}_\mathrm{Z}|=1.1$ meV is shown in Fig.~\ref{fig:TopGapky}. Here, we present $\Delta_\mathrm{top}$ at the phase $\phi=\pi$, around which the topological region is centered, as well as the nearby phase $\phi=1.2\pi$, which is situated well inside the topological region for $|\bm{E}_\mathrm{Z}|=1.1$ meV. In accordance with our earlier argument that $\bm{E}_\mathrm{Z}\perp\bm{n}_\mathrm{soc}$ should be satisfied, a topological gap $\Delta_\mathrm{top}\neq0$ for a fixed Zeeman field perpendicular to the S/N interfaces arises only around $\theta_\mathrm{soc}\approx\pm\pi/2$, that is, for dominant Dresselhaus SOC $|\alpha|\ll|\beta|$ [blue curves in Fig.~\ref{fig:TopGapky}(a); note also the axis breaks between $\theta_\mathrm{soc}=\pm0.45\pi$ and $\theta_\mathrm{soc}=\pm0.05\pi$ , in between which $\Delta_\mathrm{top}$ vanishes]. For a Zeeman field parallel to the S/N interfaces as proposed in Refs.~\onlinecite{Pientka2017:PRX,Hell2017:PRL}, $\Delta_\mathrm{top}\neq0$ only around $\theta_\mathrm{soc}\approx0$ (and $\theta_\mathrm{soc}\approx\pm\pi$; not shown here), that is, for dominant Rashba SOC $|\alpha|\gg|\beta|$ [red curves in Fig.~\ref{fig:TopGapky}(a)].

If $\bm{E}_\mathrm{Z}$ is always adjusted to the SOC, such that $\bm{E}_\mathrm{Z}\perp\bm{n}_\mathrm{soc}$, a finite topological gap $\Delta_\mathrm{top}\neq0$ can be found for any combination of SOC [Fig.~\ref{fig:TopGapky}(b)]. Even if $\bm{E}_\mathrm{Z}\perp\bm{n}_\mathrm{soc}$, however, small finite regions appear in $\bm{E}_\mathrm{Z}$-$\phi$ space where the gap $\Delta_\mathrm{top}$ closes [see also Figs.~\ref{fig:PhaseDiagram}(a)-(c)]. For $|\alpha|=|\beta|$, $\Delta_\mathrm{top}$ vanishes at $\phi=\pi$, but remains finite at phase differences $\phi\neq\pi$ [dashed blue curve in Fig.~\ref{fig:TopGapky}(b)]. On the other hand, if $\phi\neq\pi$, $\Delta_\mathrm{top}$ vanishes in a finite range of $\theta_\mathrm{soc}$ not determined by symmetry (approximately between $\theta_\mathrm{soc}=0.3\pi$ and $\theta_\mathrm{soc}=0.4\pi$ for the parameters shown here). At smaller values of $|\bm{E}_\mathrm{Z}|$, the size of these finite regions with $\Delta_\mathrm{top}=0$ typically decreases.

Hence, we have found that for $\phi=\pi$ the topological gap vanishes at $|\alpha|=|\beta|$, independent of the other parameters, while at $\phi\neq\pi$ finite regions with $\Delta_\mathrm{top}=0$ emerge for $|\alpha|\neq|\beta|$. In the following, we elucidate the nature of these gap closings to provide guidance for finding the optimal conditions for Majorana bound states to appear.

\subsubsection{Symmetry analysis and gap closings at generic momenta}\label{Sec:GenericGapClosing}

We start with the special case $\alpha=\beta$ and keep $\bm{E}_\mathrm{Z}\perp\bm{n}_\mathrm{soc}$. By performing a rotation in spin space, the Hamiltonian~(\ref{eq:BDGHam}) [or its equivalent form~(\ref{eq:BDGHamRedefined})] for $\alpha=\beta$ can be written as
\begin{align}
\tilde{H}_\mathrm{BdG}(k_y)=&\xi_p\tau_z+\sqrt{2}\alpha(\hat{p}_x-\hbar k_y)s_y\tau_z+E_Zs_x\notag\\
&+\Delta(x)[\tau_x\cos\Phi(x)-\tau_y\sin\Phi(x)],\label{hamil_0}
\end{align}
where $\xi_p$ is the kinetic energy and we have already used translational symmetry along the $y$ direction by replacing the corresponding momentum by the real parameter $k_y$. With our phase convention $\Phi(x)=(\pi-\phi)/2+\Theta(x)\phi$, a mirror reflection $x\to-x$ effects the transformation $\cos\Phi(x)\to-\cos\Phi(x)$ and $\sin\Phi(x)\to\sin\Phi(x)$.

The BdG form of the Hamiltonian dictates the presence of a particle-hole symmetry ${\cal C}\tilde{H}_\mathrm{BdG}(k_y){\cal C}^{-1}=-\tilde{H}_\mathrm{BdG}(-k_y)$ with ${\cal C}={\cal K}\tau_ys_y$ and ${\cal K}$ the complex conjugation operator. This symmetry relates positive and negative energies at opposite values of the conserved momentum $k_y$. However, there is generically no symmetry between energies of opposite sign in the spectrum of $\tilde{H}_\mathrm{BdG}(k_y)$ itself.

The situation is different for a $\pi$ junction, where an effective particle-hole symmetry ${\cal C}'={\cal K}M_x\tau_xs_y$ emerges. Here $M_x=(x\to-x)\times\i s_x$ is the mirror symmetry with respect to the $yz$ plane. This symmetry relates positive and negative energies at the {\em same} momentum,
\begin{align}
{\cal C}'\tilde{H}_\mathrm{BdG}(k_y)({\cal C}')^{-1}=-\tilde{H}_\mathrm{BdG}(k_y).\label{effective_ph_symm}
\end{align}
This property can be readily verified by using the relations ${\cal C}'\cos\Phi(x)({\cal C}')^{-1}=-\cos\Phi(x)$ and $\sin\Phi(x)=0$ for $\phi=\pi$.

\begin{figure}[t]
\centering
\includegraphics*[width=8.5cm]{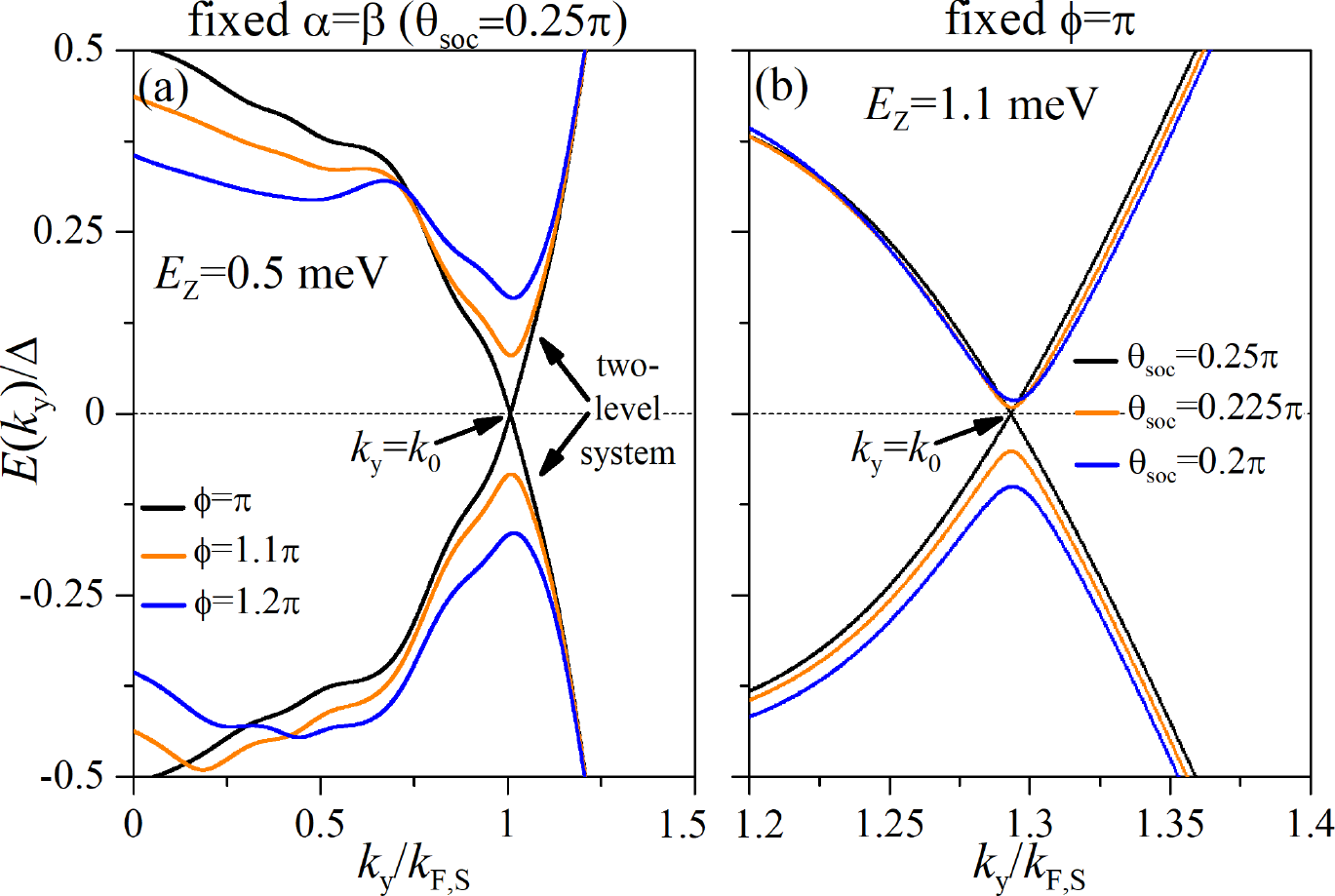}
\caption{(Color online) Andreev bound state spectrum as a function of the transverse momentum $k_y$ for (a) $\alpha=\beta$ (that is, $\theta_\mathrm{soc}=0.25\pi$) at $\phi=\pi$, $\phi=1.1\pi$, and $\phi=1.2\pi$ and for (b) a fixed $\phi=\pi$ with $\theta_\mathrm{soc}=0.25\pi$, $\theta_\mathrm{soc}=0.225\pi$, and $\theta_\mathrm{soc}=0.2\pi$. Here, the position $k_y=k_0$ of a generic gap closing and the low-energy two-level system described by Eq.~(\ref{hamil_generic}) are also indicated. In both panels, $\bm{E}_\mathrm{Z}\perp\bm{n}_\mathrm{soc}$ with $|\bm{E}_\mathrm{Z}|=0.5$ meV in panel~(a) and $|\bm{E}_\mathrm{Z}|=1.1$ meV in panel~(b). All other parameters are the same as in Fig.~\ref{fig:Finiteky}.}\label{fig:RDFiniteky}
\end{figure}

Examples for the spectra with $\alpha=\beta$ and $\bm{E}_\mathrm{Z}\perp\bm{n}_\mathrm{soc}$ [that is, the spectra of Eq.~(\ref{hamil_0})] are shown in Fig.~\ref{fig:RDFiniteky}(a). For $\phi=\pi$ and $\alpha=\beta$, the spectrum always exhibits at least one gap closing and hence the topological gap $\Delta_\mathrm{top}$ given by Eq.~(\ref{eq:TopGap}) vanishes, whereas a finite $\Delta_\mathrm{top}$ is possible for $\phi\neq\pi$. In order to understand possible gap closings, we focus on the Hilbert space spanned by the two states closest to zero energy (see Fig.~\ref{fig:RDFiniteky} for the states constituting this two-level system). Because the effective particle-hole operator has the property $({\cal C}')^2=1$, we can choose a basis in the two-dimensional low-energy subspace, in which the operator takes the form ${\cal C}'={\cal K}$. Condition~(\ref{effective_ph_symm}) requires the $2\times2$ effective Hamiltonian in this basis to take the form $\hat{H}_\mathrm{eff}(k_y)=h_y(k_y)\sigma_y$, where $\sigma_y$ is a Pauli matrix describing the two-level system and $h_y(k_y)$ is a real function of $k_y$. Because $\hat{H}_\mathrm{eff}(k_y)$ depends only on a single parameter, a gap closing can generically occur for some momentum $k_y=k_0$ without additional fine tuning (see Fig.~\ref{fig:RDFiniteky} for the position of $k_0$). To understand the closing of the gap inside the topological phase, we note that the band structure is inverted at $k_y=0$, but it remains non-inverted at large momenta. The symmetry argument above demonstrates that the bands cross (rather than avoid) each other when the momentum is tuned from zero to $k_\mathrm{F}$. Hence, the system is typically gapless at $\phi=\pi$ and $\alpha=\beta$ inside the topological phase. This is also corroborated by Figs.~\ref{fig:PhaseDiagram}(c) and~\ref{fig:TopGapky}, where $\Delta_\mathrm{top}$ always vanishes at $\phi=\pi$ for $\alpha=\beta$, independent of the other parameters. Although we have performed the above analysis for $\alpha=\beta$, the same arguments hold for the general condition $|\alpha|=|\beta|$ in junctions with phase bias along the [100] direction.

When the phase difference is tuned away from $\pi$ or $\alpha\neq\beta$, the Hamiltonian $\hat{H}_\mathrm{BdG}(k_y)$ looses the symmetry ${\cal C}'$. However, it retains an effective time-reversal symmetry ${\cal T}={\cal K}(x\to-x)\tau_zs_x$, which satisfies ${\cal T}\hat{H}_\mathrm{BdG}(k_y)({\cal T})^{-1}=\hat{H}_\mathrm{BdG}(k_y)$ (see Ref.~\onlinecite{Pientka2017:PRX}, where a similar time-reversal symmetry was discussed for the case $\beta=0$). Next, we construct an effective Hamiltonian for the two states closest to zero energy for $\phi\neq\pi$ or $\alpha\neq\beta$. Keeping the representation of the now broken symmetry ${\cal C}'$ as ${\cal C}'={\cal K}$ in the two-level system, we need to determine a corresponding representation of ${\cal T}$ in order to set up the effective Hamiltonian. The properties ${\cal T}^2=1$ and $[{\cal C}',{\cal T}]=0$ allow us to choose the low-energy basis such that ${\cal T}={\cal K}\sigma_x$. In this basis, the low-energy Hamiltonian for $\phi\neq\pi$ or $\alpha\neq\beta$ takes the general form 
\begin{align}
\hat{H}_\mathrm{eff}(k_y)=h_0(k_y)+h_x(k_y)\sigma_x+h_y(k_y)\sigma_y.\label{hamil_generic}
\end{align}

We now want to answer the question whether the spectrum is gapped or gapless when the Hamiltonian is tuned away slightly from the particle-hole invariant point (PHIP) $\alpha=\beta$ and $\phi=\pi$, where the effective particle-hole symmetry~(\ref{effective_ph_symm}) holds. The spectrum has a gap whenever $h_0^2<h_x^2+h_y^2$ and, since $h_0$ and $h_x$ are small near the PHIP, a gap closing can only occur for momenta near $k_y=k_0$, where $h_y$ vanishes. The presence of a gap therefore depends on the relative magnitude of $h_x$ and $h_0$, which are determined by the nature of the perturbation.

For $\phi\neq\pi$ but $\alpha=\beta$, the symmetry ${\cal C}'$ is broken by the term $V_1=\Delta(x)\tau_y\sin\Phi(x)$. In the absence of $V_1$, all eigenmodes of the Hamiltonian are locally eigenstates of linear combinations of $\tau_x$ and $\tau_z$. Hence, to lowest order in the perturbation $V_1\propto\tau_y$, all matrix elements are off-diagonal in the eigenmodes and thus $|h_x|>|h_0|$ near the PHIP. This means the gap closing is lifted when tuning away from $\pi$ phase difference, which is also seen in Fig.~\ref{fig:RDFiniteky}(a) and manifests itself in a finite $\Delta_\mathrm{top}$ in Fig.~\ref{fig:PhaseDiagram}(c).

The situation is somewhat different in the case $\alpha\neq\beta$ but $\phi=\pi$ [see Fig.~\ref{fig:RDFiniteky}(b)], which leads to a symmetry-breaking perturbation $V_2\propto\hat{p}_xs_x\tau_z$ to Eq.~(\ref{hamil_0}). Here, we have assumed that the Zeeman field remains perpendicular to $\bm{n}_\mathrm{soc}$ and the Hamiltonian is rotated in spin space such that $\bm{E}_\mathrm{Z}\parallel\hat{x}$. In the case $\alpha k_\mathrm{F,N}\gg|\bm{E}_\mathrm{Z}|$, the relevant states near the Fermi surface are mostly spin polarized along the $y$ direction. Hence, the perturbation $V_2\propto s_x$ will be mostly off-diagonal in the eigenbasis, which results in $|h_x|>|h_0|$. In the opposite case $\alpha k_\mathrm{F,N}\ll|\bm{E}_\mathrm{Z}|$, the spins on the Fermi surface are mostly polarized by the Zeeman field along the $x$ direction and the perturbation $V_2$ leads to a uniform energy shift of all low-energy states, that is, $|h_0|\gg|h_x|$. This situation is illustrated by Fig.~\ref{fig:RDFiniteky}(b), which shows the low-energy Andreev spectrum in the vicinity of $k_y=k_0$ for relatively large $|\bm{E}_\mathrm{Z}|$. We conclude, that $V_2$ lifts the gap closing for strong SOC, $\alpha k_\mathrm{F,N}\gg|\bm{E}_\mathrm{Z}|$, whereas in the opposite limit the system is gapless even away from $\alpha=\beta$. Such a trend can also be seen in Fig.~\ref{fig:PhaseDiagram}(b), where the size of the gapless regions increases with $|\bm{E}_\mathrm{Z}|$.

Finally, when both perturbations $V_1$ and $V_2$ are present, additional gap closings can occur away from the PHIP. For a fixed value of $\alpha\neq\beta$, one needs to tune two parameters to encounter a gap closing of Eq.~(\ref{hamil_generic}), for example, $h_x$ and $h_y$. This can be achieved if both $k_y$ and $\phi$ are tuned. Because of the nonzero value of $h_0$, the system typically remains gapless over a range of parameters. This explains why the gapless line at $\phi=\pi$ in Fig.~\ref{fig:PhaseDiagram}(c) shifts to other phase values [and becomes an extended region as shown, for example, in Fig.~\ref{fig:TopGapky}(b)] when $\alpha$ and $\beta$ are made unequal.

\section{Conditions for the topological phase in Josephson junctions with superconducting phase bias along the [110] direction}\label{Sec:TP110}

Section~\ref{Sec:TP100} has been devoted to the appearance of a topological phase in Josephson junctions with strong SOC and superconducting phase bias along the crystallographic [100] direction. Josephson junctions based on a quantum well grown in the [001] direction can, however, also be set up such that the phase bias is along another crystallographic direction.\cite{Hart2017:NP} Such a different setup of the Josephson junction also affects the interplay between Rashba and Dresselhaus SOC. As the most pronounced example of how the phase diagram can be affected, we briefly discuss the emergence of a topological phase in junctions with phase bias along the [110] direction [see Fig.~\ref{fig:Scheme}(d) for an example of the Fermi contours in such a junction]. If we denote the [110] direction as the $x$ direction and the [1$\bar{1}$0] direction as the $y$ direction, the BdG Hamiltonian of this system is
\begin{equation}\label{eq:BDGHam_o110}
\begin{array}{l}
\hat{H}_\mathrm{BdG}=\left[\frac{\hat{p}_x^2+\hat{p}_y^2}{2m}+\frac{\alpha}{\hbar}\left(s_y\hat{p}_x-s_x\hat{p}_y\right)-\frac{\beta}{\hbar}\left(s_y\hat{p}_x+s_x\hat{p}_y\right)\right.\\
\quad\quad\quad\left.+\frac{m\alpha^2}{2\hbar^2}+\frac{m\beta^2}{2\hbar^2}-\mu\right]\tau_z+\left(V_0\tau_z-\bm{E}_\mathrm{Z}\cdot\bm{s}\right)h(x)\\
\quad\quad\quad+\Delta(x)\left[\tau_x\cos\Phi(x)-\tau_y\sin\Phi(x)\right],
\end{array}
\end{equation}
where $h(x)$, $\Delta(x)$, and $\Phi(x)$ are defined as for Eq.~(\ref{eq:BDGHam}) above. Equation~(\ref{eq:BDGHam_o110}) is the same as Eq.~(\ref{eq:BDGHam}) with the exception that the Dresselhaus term has been rotated. Here, we emphasize that the growth direction (that is, the $z$ direction) of the quantum well described by Eq.~(\ref{eq:BDGHam_o110}) is still the [001] direction.

\begin{figure}[t]
\centering
\includegraphics*[width=8.5cm]{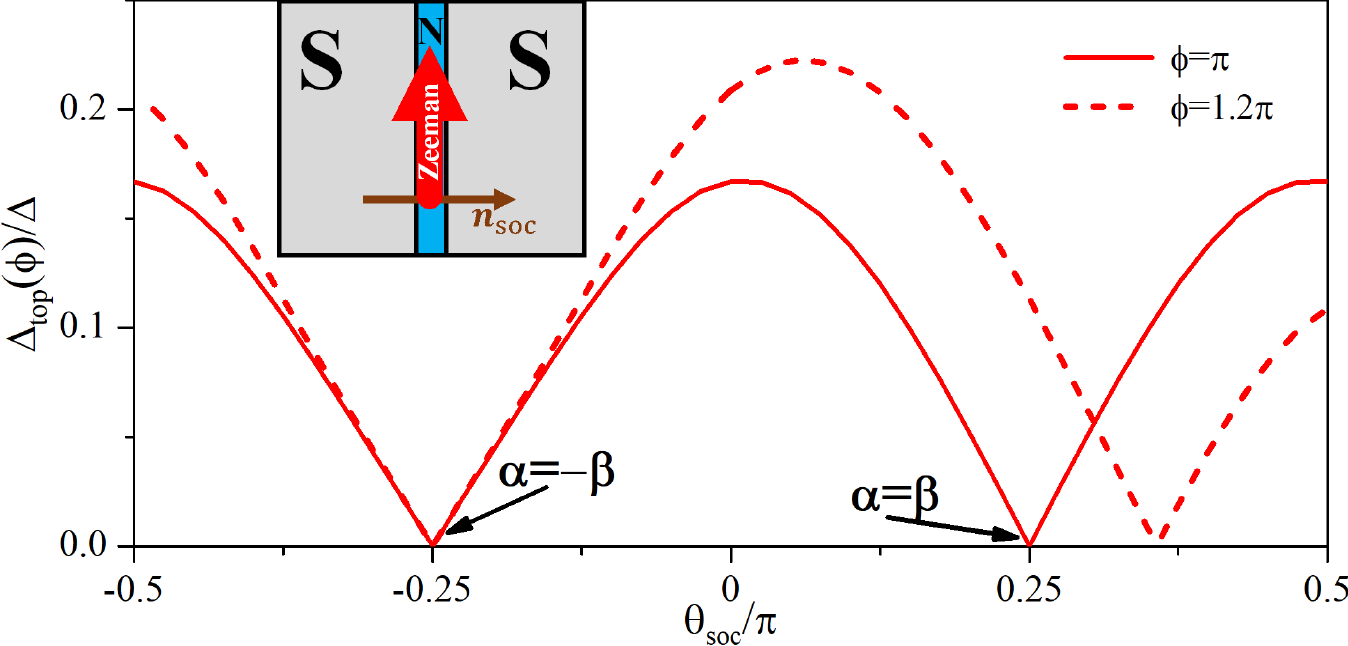}
\caption{(Color online) Dependence of the topological gap $\Delta_\mathrm{top}$ at $\phi=\pi$ (solid curve) and $\phi=1.2\pi$ (dashed curve) on $\theta_\mathrm{soc}$ for a Josephson junction with a phase bias along the [110] direction. Here, $\bm{E}_\mathrm{Z}=|\bm{E}_\mathrm{Z}|\bm{e}_y$ is kept constant with $|\bm{E}_\mathrm{Z}|=1.1$ meV. The other parameters are the same as in Fig.~\ref{fig:Finiteky}. The inset shows a scheme of the setup.}\label{fig:TopGapkyO110}
\end{figure}

Analogously to Sec.~\ref{Sec:Finiteky}, we can introduce SO fields, which now read as
\begin{equation}\label{eq:hnSOCo110}
\bm{h}_\mathrm{soc}=(\alpha-\beta)\bm{e}_y\quad\text{and}\quad\bm{n}_\mathrm{soc}=(\alpha+\beta)\bm{e}_x,
\end{equation}
and recast Eq.~(\ref{eq:BDGHam_o110}) into Eq.~(\ref{eq:BDGHamRedefined}). Then, we can use the arguments presented in Sec.~\ref{Sec:TP100} to argue for the appearance of a topological phase around $\phi=\pi$ and $|\bm{E}_\mathrm{Z}|=E_\mathrm{T}$ in $\bm{E}_\mathrm{Z}$-$\phi$-space. The only change compared to Sec.~\ref{Sec:TP100} now is that $\bm{n}_\mathrm{soc}$ is always perpendicular to the S/N interfaces. Hence, $\bm{E}_\mathrm{Z}\perp\bm{n}_\mathrm{soc}$ requires Zeeman fields parallel to the S/N interfaces, $\bm{E}_\mathrm{Z}=|\bm{E}_\mathrm{Z}|\bm{e}_y$, regardless of the combination of SOC.

The specific form of $\bm{n}_\mathrm{soc}$, moreover, means that $\bm{n}_\mathrm{soc}$ vanishes completely for $\alpha=-\beta$. Hence, no gapped spectrum arises for an infinitely long junction in the $y$ direction because in this case the Hamiltonian contains no terms linear in $\hat{p}_y$. Since such linear terms are necessary for the formation of localized 1D edge states along the $y$ direction, no Majorana bound states and topological phase emerge for $\alpha=-\beta$.

In Fig.~\ref{fig:TopGapkyO110}, we again show the dependence of $\Delta_\mathrm{top}$ (for $\alpha\geq0$) on $\theta_\mathrm{soc}$ for the same parameters as in Fig.~\ref{fig:Finiteky}, but now with phase bias along the [110] direction. Similar to Fig.~\ref{fig:TopGapky}, $\Delta_\mathrm{top}$ at $\phi=\pi$ is typically larger if either Rashba or Dresselhaus SOC is dominant and $\Delta_\mathrm{top}$ tends to zero for $|\alpha|=|\beta|$. Furthermore, we again emphasize that for $\alpha=-\beta$, no Majorana bound states can arise and consequently $\Delta_\mathrm{top}=0$ for any phase difference $\phi$. For $\alpha=\beta$ in contrast, the Hamiltonian~(\ref{eq:BDGHam_o110}) still contains $\hat{p}_y$-linear terms, necessary for the formation of edge states along the $y$ direction. This situation is similar to the case $\alpha=\beta$ discussed in Sec.~\ref{Sec:TP100} for junctions along the [100] direction, where the high-symmetry point at $\alpha=\beta$ and $\phi=\pi$ exhibits a gap closing: Although $\Delta_\mathrm{top}=0$ at $\phi=\pi$ in this case, a finite topological gap arises at phase differences $\phi\neq\pi$ and well-localized Majorana bound states can form.

Concluding the discussion of the S/N/S junctions with infinite extensions in $y$ direction from Secs.~\ref{Sec:TP100} and~\ref{Sec:TP110}, we have found that an extended topological phase can appear for an arbitrary combination of Rashba and Dresselhaus SOC if $\bm{E}_\mathrm{Z}\perp\bm{n}_\mathrm{soc}$. This topological phase emerges in a diamond centered around $\phi=\pi$ and $|\bm{E}_\mathrm{Z}|=E_\mathrm{T}$ in $\bm{E}_\mathrm{Z}$-$\phi$-space. Inside this diamond, there can, however, be lines or regions with $\Delta_\mathrm{top}=0$. An example of this is the line with $\Delta_\mathrm{top}=0$ due to the gap closing at $\phi=\pi$ for $|\alpha|=|\beta|$ discussed in Sec.~\ref{Sec:GenericGapClosing}. Moreover, the topological gap protecting this phase is maximal if one kind of SOC is dominant. In the following, we will now explicitly look into the appearance of Majorana bound states and potential experimental signatures of these states.

\section{Majorana end states in Josephson junctions}\label{Sec:MBS}

\begin{figure}[t]
\centering
\includegraphics*[width=8.5cm]{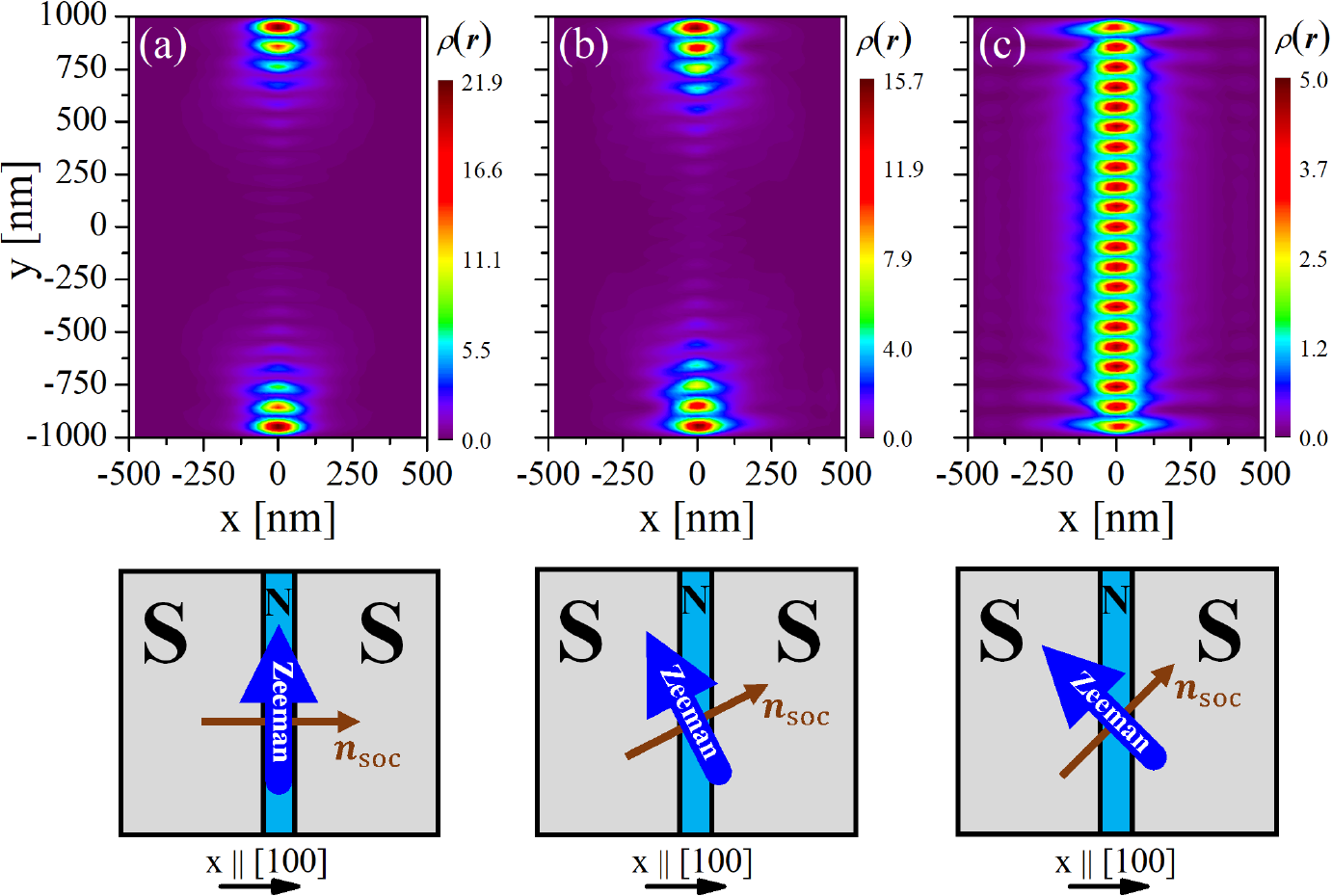}
\caption{(Color online) Probability densities $\rho(\bm{r})=|\psi(\bm{r})|^2$ (in $\mu$m$^{-2}$) of Majorana bound states in a finite system with $W=100$ nm, $W_\mathrm{S}=450$ nm, and finite length $L=2$ $\mu$m. Here, a superconducting phase difference $\phi=\pi$ and several different configurations of SOC and $\bm{E}_\mathrm{Z}$ are shown: (a) $\theta_\mathrm{soc}=0$ ($\alpha=16$ meVnm, $\beta=0$), $E_{\mathrm{Z},x}=0$, $E_{\mathrm{Z},y}=0.5$ meV, (b) $\theta_\mathrm{soc}=0.15\pi$ ($\alpha\approx14.3$ meVnm, $\beta\approx7.3$ meVnm), $E_{\mathrm{Z},x}=-0.21$ meV, $E_{\mathrm{Z},y}=0.41$ meV, (c) $\theta_\mathrm{soc}=0.25\pi$ ($\alpha=\beta\approx11.3$ meVnm), $E_{\mathrm{Z},x}=-E_{\mathrm{Z},y}=-0.35$ meV. Schemes of the configurations investigated are shown below the density plots. In all panels, $|\bm{E}_\mathrm{Z}|=0.5$ meV, $\lambda_\mathrm{soc}=16$ meVnm, $m=0.038m_0$, $\mu_\mathrm{S}=1$ meV, $\mu_\mathrm{N}=0.7$ meV, and $\Delta=250$ $\mu$eV. The $x$ direction is chosen along the crystallographic [100] direction.}\label{fig:MBS}
\end{figure}

Having discussed the 1D band structure of a Josephson junction which is infinite in the $y$ direction, we now verify our predictions of the appearance of Majorana bound states. We do so by considering a system that is also confined along the $y$ direction with a finite length $L=2$ $\mu$m. The eigenspectrum and eigenstates of this system, described by Eqs.~(\ref{eq:BDGHam}) or~(\ref{eq:BDGHam_o110}), are obtained by a 2D finite-difference method with hard-wall boundary conditions along the $x$ and $y$ directions. As expected from the previous discussion in Secs.~\ref{Sec:TP100} and~\ref{Sec:TP110}, no localized Majorana bound states can be found if the Zeeman field $\bm{E}_\mathrm{Z}$ deviates significantly from $\bm{E}_\mathrm{Z}\perp\bm{n}_\mathrm{soc}$. If $\bm{E}_\mathrm{Z}\perp\bm{n}_\mathrm{soc}$, on the other hand, Majorana bound states can be found, examples of which are shown in Fig.~\ref{fig:MBS} for different configurations of SOC.

We show results for a phase difference along the [100] direction, but remark that the results for a geometry with phase bias along the [110] direction would look qualitatively similar. The phase difference is chosen as $\phi=\pi$. Figure~\ref{fig:MBS}(a) shows Majorana bound states appearing in a system with only Rashba SOC, $\theta_\mathrm{soc}=0$. The corresponding phase diagram of this system can be found in Fig.~\ref{fig:PhaseDiagram}(a).

The situation in Fig.~\ref{fig:MBS}(a) is also representative for typical configurations with mixed Rashba and Dresselhaus SOC where $\bm{E}_\mathrm{Z}\perp\bm{n}_\mathrm{soc}$, as illustrated by Fig.~\ref{fig:MBS}(b): Again, Majorana bound states appear at the ends of the N region. Compared to the Fig.~\ref{fig:MBS}(a), the Majorana bound states are slightly less localized. This difference can also be understood from the phase diagrams in Figs.~\ref{fig:PhaseDiagram}(a) and~(b), where the topological gap $\Delta_\mathrm{top}$ is typically somewhat smaller for $\theta_\mathrm{soc}=0.15\pi$ than for $\theta_\mathrm{soc}=0$. Consequently, Majorana bound states are expected to be less localized for $\theta_\mathrm{soc}=0.15\pi$ [Fig.~\ref{fig:MBS}(b)] because their localization lengths are proportional to $1/\Delta_\mathrm{top}$. Additionally, we note that there is also a slight asymmetry with respect to $x\to-x$ due to the tilted Zeeman field in Fig.~\ref{fig:MBS}(b).

Although Figs.~\ref{fig:MBS}(a) and~(b) display typical situations with well-pronounced Majorana bound states if $\bm{E}_\mathrm{Z}\perp\bm{n}_\mathrm{soc}$, the situation is different for $\alpha=\beta$ and $\phi=\pi$. This case is representative for situations, where $\Delta_\mathrm{top}\to0$ and the low-energy states become delocalized across the junction. An example of this is exhibited in Fig.~\ref{fig:MBS}(c). Now the lowest-energy states are extending far into the bulk of the N region and are indistinguishable from conventional Andreev bound states. Note, however, that for $\alpha=\beta$ and phase differences $\phi\neq\pi$ localized Majorana bound states like those shown in Figs.~\ref{fig:MBS}(a) and~(b) can emerge. Likewise, for $|\alpha|\neq|\beta|$ and phase differences $\phi$, where $\Delta_\mathrm{top}\to0$, the low-energy states are qualitatively similar to the state shown in Fig.~\ref{fig:MBS}(c) in that they are completely delocalized in the N region.

\begin{figure*}[t]
\centering
\includegraphics*[width=17.0cm]{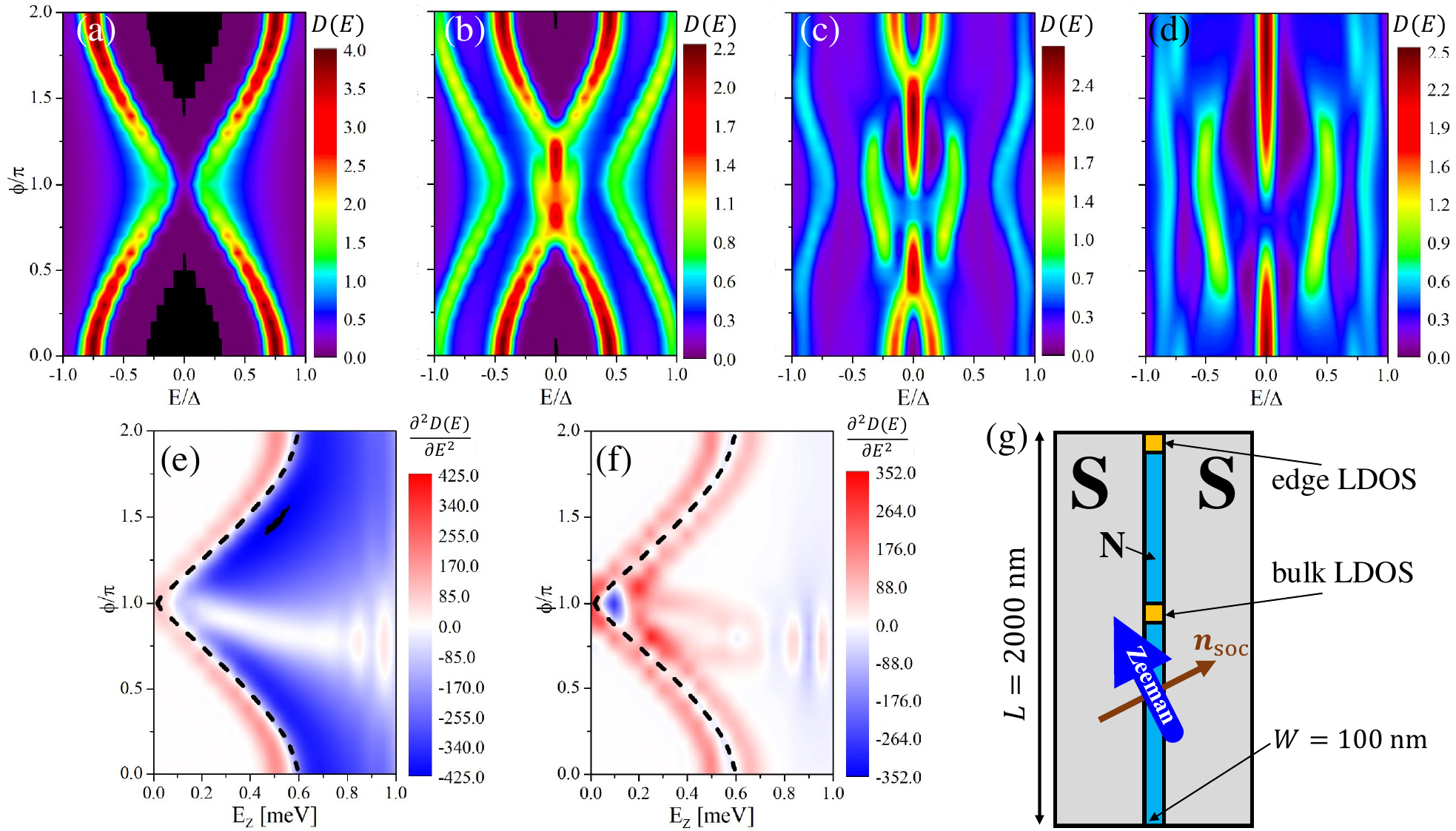}
\caption{(Color online) Local density of states $D(E)$ for a Josephson junction with $W=100$ nm, $W_\mathrm{S}=450$ nm, $L=2$ $\mu$m, $m=0.038m_0$, $\mu_\mathrm{S}=1$ meV, $\mu_\mathrm{N}=0.7$ meV, $\Delta=250$ $\mu$eV, $\theta_\mathrm{soc}=0.15\pi$ ($\alpha=14.3$ meVnm, $\beta=7.3$ meVnm), and $\bm{E}_\mathrm{Z}\perp\bm{n}_\mathrm{soc}$. (a)-(d) $D(E)$ (in a.u.) measured at the top edge of the N region for different Zeeman terms: (a) $|\bm{E}_\mathrm{Z}|=0$, (b) $|\bm{E}_\mathrm{Z}|=0.23$ meV, (c) $|\bm{E}_\mathrm{Z}|=0.46$ meV, (d) $|\bm{E}_\mathrm{Z}|=0.69$ meV. (e) and~(f) $\partial^2D/\partial E^2|_{E=0}$ (in a.u.) measured (e) at the top and (f) in the center of the N region as a function of $|\bm{E}_\mathrm{Z}|$ and $\phi$. The dashed black lines denote the phase boundaries obtained from Eq.~(\ref{eq:PhaseBoundary}) for $L\to\infty$. (g) Scheme of the setup investigated with the position of the edge and bulk probes indicated.}\label{fig:LDOSRotatedField}
\end{figure*}

\section{Signatures of Majorana bound states in the local density of states}\label{Sec:ExpSig}

In Secs.~\ref{Sec:TP100} and~\ref{Sec:MBS}, it was shown that Majorana bound states appear in phase-controlled Josephson junctions as long as the following conditions are satisfied: First, the magnetic field must be oriented in such a way that it is perpendicular to the spin-orbit field $\bm{n}_\mathrm{soc}$, which describes propagation parallel to the S/N interfaces. Second, the Zeeman energy $|\bm{E}_\mathrm{Z}|$ must be large enough to overcome normal reflection. Now, we discuss possible experimentally observable signatures of these Majorana bound states. A potential venue is to use tunneling spectroscopy to probe the local density of states (LDOS), which to lowest order is proportional to the $\d I/\d V$ characteristics of the tunneling current between the tunneling probe and the junction.\cite{Ren2019:N} The position of the tunneling probe then determines at which point in real space the LDOS is measured.

In order to describe this situation, we again study a finite system in $x$ and $y$ directions with widths $W$, $W_\mathrm{S}$, and $L$. Invoking hard-wall boundary conditions, the eigenstates $\psi_n(\bm{r})$ of Eq.~(\ref{eq:BDG}) and their corresponding energies $E^{(n)}$ are obtained by the finite-difference method, similar to Sec.~\ref{Sec:MBS}. Here, $n$ labels the different eigenstates. Next, we compute the LDOS $D(E)$ for the area $W_\mathrm{p}\times L_\mathrm{p}$ covered by the tunneling probe,
\begin{equation}\label{eq:LDOS}
D(E)=\sum\limits_n\int\limits_{W_\mathrm{p}\times L_\mathrm{p}}\d^2r\left|\psi_n(\bm{r})\right|^2\delta_\Gamma(E-E^{(n)}),
\end{equation}
where $W_\mathrm{p}$ and $L_\mathrm{p}$ denote the extensions of the tunneling probe in the $x$ and $y$ directions, respectively. We model $\delta_\Gamma(E-E^{(n)})$ by a Gaussian with broadening $\Gamma$, which is chosen as $\Gamma=0.05\Delta$ in the following.

In the LDOS measured at the ends $y=\pm L/2$ of the N region, Majorana bound states should manifest themselves as zero-energy peaks. At this point, it is important to recall that Majorana bound states are, however, not the only possible source for zero-energy peaks. Such peaks can also arise from conventional bulk Andreev bound states at or close to $E=0$ with sufficient weight at the edge. Here, the term bulk Andreev bound state is employed in the sense that these states remain if the system does not have any boundaries in the $y$ direction. In fact, it is the bulk Andreev bound states around $E=0$ that describe the boundary $\phi_c$ of the topological phase, discussed previously for the infinite system with Eq.~(\ref{eq:PhaseBoundary}). Hence, when studying the edge LDOS around $E=0$, the competition between Majorana and bulk Andreev bound states should be kept in mind.

Figures~\ref{fig:LDOSRotatedField}(a)-(d) show the LDOS measured at the upper end of the N region for the narrow junction with mixed SOC studied in Fig.~\ref{fig:MBS}(b) and different strengths of the Zeeman field $\bm{E}_\mathrm{Z}\perp\bm{n}_\mathrm{soc}$. The area of integration in Eq.~(\ref{eq:LDOS}) is $W_\mathrm{p}\times L_\mathrm{p}=100$nm$\times100$nm and indicated in Fig.~\ref{fig:LDOSRotatedField}(g). While there are no zero-energy peaks at $|\bm{E}_\mathrm{Z}|=0$ due to normal reflection, well-localized Majorana bound states with pronounced zero-energy peaks in the LDOS emerge in Figs.~\ref{fig:LDOSRotatedField}(b)-(d). The region in $\phi$-space in which zero-energy peaks appear increases with increasing $|\bm{E}_\mathrm{Z}|$, consistent with the phase diagram presented in Fig.~\ref{fig:PhaseDiagram}(b). Moreover, despite the appearance of such a zero-energy peak at $\phi=\pi$ due to Majorana bound states in Figs.~\ref{fig:LDOSRotatedField}(b)-(d), this peak can be even more pronounced at certain $\phi\neq0$, a behavior which also follows the behavior of $\Delta_\mathrm{top}$ in Fig.~\ref{fig:PhaseDiagram}(b).

The comparison with Fig.~\ref{fig:PhaseDiagram}(b) can be made more quantitative by considering the curvature of the edge LDOS with respect to energy, $\partial^2D/\partial E^2$, at $E=0$. Then, zero-energy peaks are described by $\partial^2D/\partial E^2|_{E=0}<0$ and dips by $\partial^2D/\partial E^2|_{E=0}>0$. Plotting $\partial^2D/\partial E^2|_{E=0}$ as a function of the Zeeman energy $|\bm{E}_\mathrm{Z}|$ and the phase difference $\phi$ allows us to reconstruct the phase diagram,\cite{Ren2019:N} as illustrated by Fig.~\ref{fig:LDOSRotatedField}(e): Here, $\partial^2D/\partial E^2|_{E=0}$ at the upper end of the N region is shown. Inside the topological phase, obtained from Eq.~(\ref{eq:PhaseBoundary}) for $L\to\infty$, we find pronounced zero-energy peaks ($\partial^2D/\partial E^2|_{E=0}<0$) due to the Majorana bound states. A comparison of Fig.~\ref{fig:LDOSRotatedField}(e) with Fig.~\ref{fig:PhaseDiagram}(b) for $|\bm{E}_\mathrm{Z}|\leq1$ meV reveals that a large absolute value of $\partial^2D/\partial E^2|_{E=0}<0$ in the finite system with Majorana bound states corresponds to a large topological gap $\Delta_\mathrm{top}$ in the infinite system, as expected from the discussion above.

Hence, the edge LDOS in narrow junctions can be used to reconstruct the topological phase diagram. The contrast between the edge and bulk LDOS, whose curvature is shown in Fig.~\ref{fig:LDOSRotatedField}(f), allows one to identify the signatures in the edge LDOS as emerging from the Majorana end states. In the bulk LDOS, measured for an area of $W_\mathrm{p}\times L_\mathrm{p}=100$nm$\times100$nm in the center of the N region, we typically do not find pronounced zero-energy peaks. Away from the phase boundary, which is effectively given by the bulk Andreev bound states close to $E=0$, there are no other bulk states around $E=0$ and hence $\partial^2D/\partial E^2|_{E=0}\approx0$.

\begin{figure}[t]
\centering
\includegraphics*[width=8.5cm]{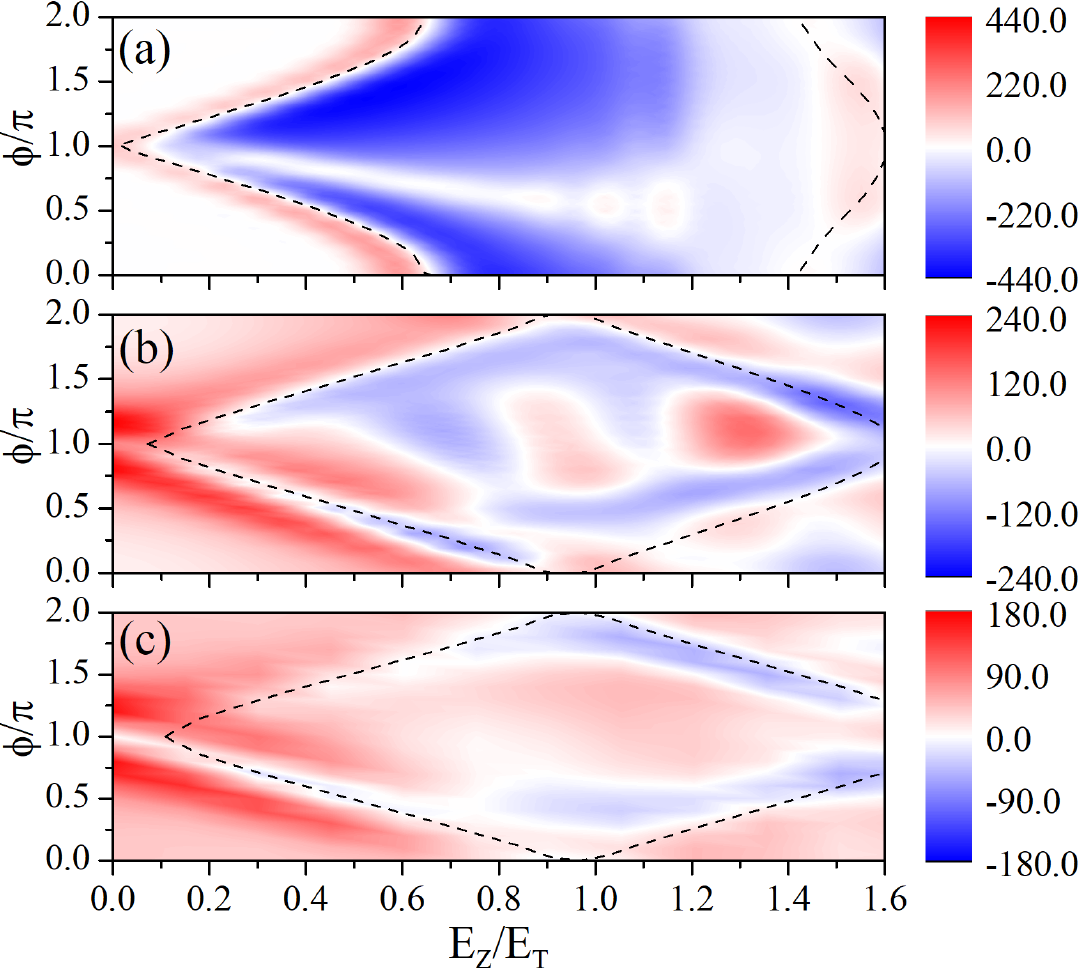}
\caption{(Color online) Curvature of the edge LDOS, $\partial^2D/\partial E^2|_{E=0}$ (in a.u.), as a function of the Zeeman field $|\bm{E}_\mathrm{Z}|$ and the superconducting phase difference $\phi$ for different widths $W$ of the N region: (a) $W=100$ nm, (b) $W=300$ nm, and (c) $W=500$ nm. The other parameters are chosen as in Fig.~\ref{fig:MBS}(a): $W_\mathrm{S}=450$ nm, $L=2$ $\mu$m, $m=0.038m_0$, $\mu_\mathrm{S}=1$ meV, $\mu_\mathrm{N}=0.7$ meV, $\Delta=250$ $\mu$eV, $\lambda_\mathrm{soc}=16$ meVnm, $\theta_\mathrm{soc}=0$, and $\bm{E}_\mathrm{Z}=|\bm{E}_\mathrm{Z}|\bm{e}_y$. In all panels, the dashed black lines denote the phase boundaries obtained from Eq.~(\ref{eq:PhaseBoundary}) for $L\to\infty$.}\label{fig:LDOSwidths}
\end{figure}

Until now, we have mainly discussed narrow junctions with $\Delta<E_\mathrm{T}$, where sizable topological gaps $\Delta_\mathrm{top}$ can arise for $L\to\infty$. As mentioned in Sec.~\ref{Sec:TopGap}, going to wider junctions reduces $\Delta_\mathrm{top}$ which has an upper boundary of the order of $\hbar^2/mW^2$. Then, the Majorana bound states are less localized and, as a consequence, the Majorana features discussed in Fig.~\ref{fig:LDOSRotatedField} become much less prominent.

This is illustrated by Fig.~\ref{fig:LDOSwidths}, which again shows the curvature of the LDOS measured for an area of $W_\mathrm{p}\times L_\mathrm{p}=100$nm$\times100$nm centered around $x=0$ at the upper end of the N region for the parameters of Fig.~\ref{fig:MBS}(a) and different widths $W$ of the N region: $W=100$ nm (corresponding to $E_\mathrm{T}=832$ $\mu$eV), $W=300$ nm (corresponding to $E_\mathrm{T}=277$ $\mu$eV), and $W=500$ nm (corresponding to $E_\mathrm{T}=166$ $\mu$eV). To better compare these junctions with different $W$, the Zeeman energies $E_\mathrm{Z}$ are measured in units of the Thouless energy $E_\mathrm{T}$. Here, the system parameters are those of a system with phase bias along the [100] direction, pure Rashba SOC (or pure Dresselhaus), and $\bm{E}_\mathrm{Z}\perp\bm{n}_\mathrm{soc}$, but systems with mixed Rashba and Dresselhaus SOC produce similar features.

Similar to Fig.~\ref{fig:LDOSRotatedField}(e), the edge LDOS of the narrow junction with $W=100$ nm exhibits pronounced zero-energy peaks [blue areas in Fig.~\ref{fig:LDOSwidths}(a)] in almost the entire topological phase. As the width is increased to $W=300$ nm, the Majorana bound states become less localized and their contribution to the edge LDOS is washed out due to nearby Andreev bound states and broadening in energy. Consequently, pronounced zero-energy peaks in the edge LDOS appear only in some regions of the topological phase, where still relatively large values of $\Delta_\mathrm{top}$ allow for well-separated Majorana bound states, and at the phase boundaries, where the zero-energy peak arises from the Andreev bound states close to $E=0$.

For an even wider junction with $W=500$ nm, zero-energy peaks in the edge LDOS appear only at or close to the phase boundaries, as $\Delta_\mathrm{top}$ and the separation in energy between Majorana and Andreev bound states decrease even further. Then, the Majorana bound state is completely delocalized and cannot be distinguished from bulk states and the Majorana features in the LDOS are completely washed out. Now, the zero-energy peaks in the edge LDOS arise predominantly from the bulk Andreev bound states at $E\approx0$. These bulk states at $E\approx0$ in turn provide the boundary between the trivial and topological superconducting phases. Which sections of the phase boundaries feature more prominent peaks depends on the actual system parameters, such as the chemical potentials or the strength of SOC. For the parameters chosen in Fig.~\ref{fig:LDOSwidths}(c), the peaks at the phase boundaries for $E_\mathrm{Z}>E_\mathrm{T}$ are more pronounced than the corresponding peaks for $E_\mathrm{Z}<E_\mathrm{T}$.

Due to the Majorana bound states being no longer very localized and spread over the entire length $L$ of the N region, the edge and bulk LDOS in wide junctions are qualitatively similar and dominated by the bulk Andreev bound states. One should keep in mind, however, that the boundaries indicate a topological transition to a regime where Majorana bound states could in general form: If the length $L$ of the junction was increased while all other parameters were left unchanged, localized Majorana bound states would eventually appear. In this case, the Majorana bound states will dominate the edge LDOS and a zero-energy peak should appear for a wide range of Zeeman energies $E_\mathrm{Z}$ and superconducting phase differences $\phi$.

From Fig.~\ref{fig:LDOSwidths} and the above discussion, it follows that the aspect ratio between $L$ and $W$ plays an important role in what can be observed experimentally: Whereas measurements of the edge LDOS in wide junctions can be used to probe (parts of) the boundaries of the topological phase diagram, these measurements can in narrow junctions also provide information about the full topological phase diagram, differentiating between regions with large or small topological gaps. Recently, we have invoked this argument to explain experiments in wide HgTe/thin-film Al-based Josephson junctions.\cite{Ren2019:N}

\section{Tuning and testing topological superconductivity}\label{Sec:Tuning}

Until now, we have studied the topological phase diagram in detail in Secs.~\ref{Sec:TP100} and~\ref{Sec:TP110} and have explained how this phase diagram can be reconstructed experimentally in narrow junctions in Sec.~\ref{Sec:ExpSig}. In this section, which constitutes a central part of our manuscript, we finally discuss how our findings could be utilized to tune and test topological superconductivity and Majorana bound states in narrow, phase-controlled Josephson junctions with Rashba and Dresselhaus SOC.

First of all, Figs.~\ref{fig:PhaseDiagram}, \ref{fig:TopGapky} and~\ref{fig:TopGapkyO110} demonstrate that by tuning the ratio between Rashba and Dresselhaus SOC, $\alpha/\beta$, and orienting the in-plane magnetic field appropriately, one can significantly alter the topological phase diagram. This is especially true for phase differences around $\phi=\pi$: A finite topological gap $\Delta_\mathrm{top}$ and corresponding Majorana bound states occur at $\phi=\pi$ if either Rashba or Dresselhaus SOC are dominant, but for $\alpha=\beta$ the gap vanishes at $\phi=\pi$. Hence, when searching for Majorana bound states in materials with comparable Rashba and Dresselhaus SOC, one has to tune the phase difference away from $\phi=\pi$. Our predictions for $\alpha=\beta$ could directly be tested in materials with relatively large Dresselhaus SOC and $g$-factors, such as InAs or InSb quantum wells,\cite{Fabian2007:APS} by an independent tuning of $\alpha$ compared to a fixed $\beta$. Modulating the asymmetry in the confinement potential of electrons by gating allows one to vary $\alpha$. Such a tuning of $\alpha$ versus $\beta$ has, for example, been employed in GaAs quantum wells with [001] growth direction to observe a persistent spin helix at $\alpha=\beta$.\cite{Bernevig2006:PRL,Koralek2009:N}\footnote{In quantum wells grown along the crystallographic [001] direction, SU(2) symmetry is preserved for $\alpha=\beta$ in the absence of a magnetic field and cubic Dresselhaus terms. The corresponding conserved quantity is a helical spin density wave, termed the persistent spin helix, along a 'magic' wave vector. At this wave vector the spin lifetime is infinite if SU(2) symmetry is not broken as discussed in Ref.~\onlinecite{Bernevig2006:PRL}.} Its small $g$-factor, however, makes GaAs an unsuitable candidate to observe topological superconductivity, unlike InAs\cite{Mayer2019:arxiv} or InSb.\cite{Ke2019:arxiv} Keeping the phase difference at $\phi=\pi$ and adjusting $\alpha$ versus $\beta$, one should observe a disappearance and reemergence of Majorana bound states or corresponding zero-energy peaks in the $\d I/\d V$ characteristics of narrow junctions. This should be applicable for a wide range of Zeeman energies as long as $|\bm{E}_\mathrm{Z}|\ll\lambda_\mathrm{soc}k_\mathrm{F,N}$.

\begin{figure}[t]
\centering
\includegraphics*[width=8.5cm]{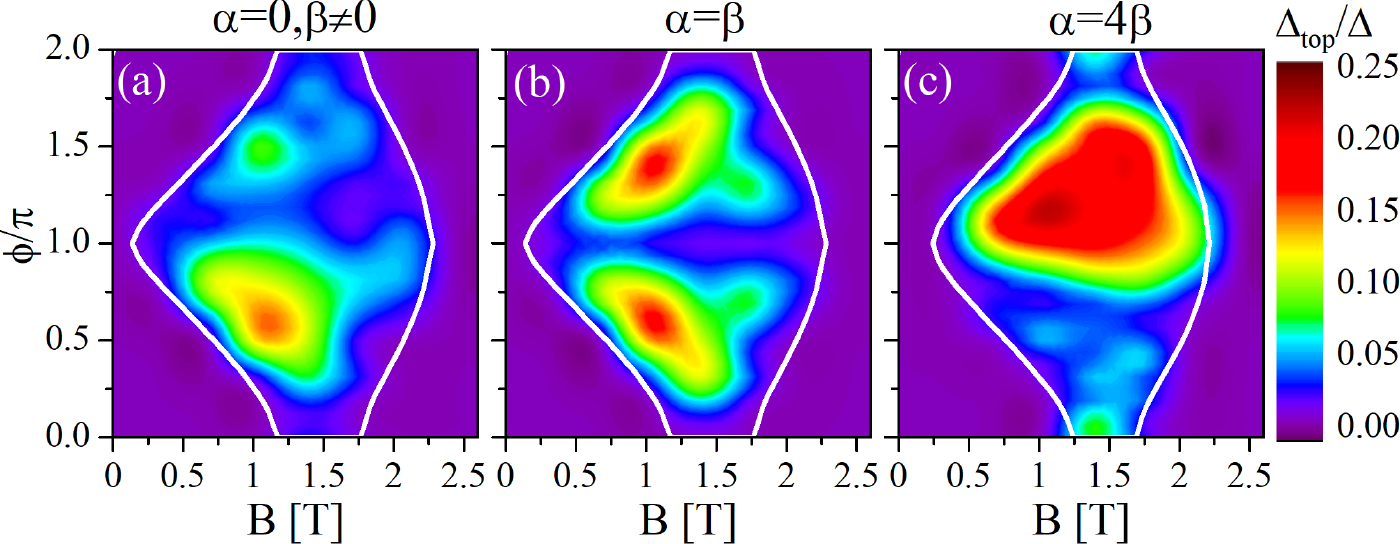}
\caption{(Color online) Topological gap $\Delta_\mathrm{top}$ for an InAs quantum well with fixed $\beta=4$ meVnm and different strengths of Rashba SOC: (a) $\alpha=0$, (b) $\alpha=\beta=4$ meV, and (c) $\alpha=16$ meVnm. The junction is set up with phase bias along the [110] direction and a magnetic field parallel to the S/N interfaces, $\bm{B}=B\bm{e}_y$. Here, $W=200$ nm, $W_\mathrm{S}=500$ nm, $m=0.026m_0$, $\mu_\mathrm{S}=1$ meV, $\mu_\mathrm{N}=0.8$ meV, $g=10$ and $\Delta=150$ $\mu$eV, corresponding to a InAs/Al heterostructure.\cite{Fornieri2019:N} For $g=10$, a magnetic field $B=1$ T corresponds to a Zeeman energy $E_\mathrm{Z}=g\mu_BB/2\approx0.29$ meV.}\label{fig:PhaseDiagramInAs}
\end{figure}

Figure~\ref{fig:PhaseDiagramInAs} shows an example for tuning topological superconductivity at $\phi=\pi$. Here, we present the phase diagrams for parameters corresponding to an InAs/Al heterostructure\cite{Fornieri2019:N} with fixed $\beta=4$ meVnm as Rashba SOC is increased from $\alpha=0$ to $\alpha=16$ meVnm. The junction is set up with phase bias along the [110] direction and a magnetic field $\bm{B}=B\bm{e}_y$ is applied parallel to the S/N interfaces. Setting up the Josephson junction with [100] phase bias instead and adjusting $\bm{B}$ appropriately to each combination of $\alpha$ and $\beta$ yields results that are qualitatively very similar to those in Fig.~\ref{fig:PhaseDiagramInAs}. The magnetic field $\bm{B}$ gives rise to a Zeeman field $\bm{E}_\mathrm{Z}=g\mu_B\bm{B}/2$ with the Bohr magneton $\mu_B$ and a $g$-factor of $g=10$. Similar to Fig.~\ref{fig:PhaseDiagram}, Fig.~\ref{fig:PhaseDiagramInAs} shows the topological gap $\Delta_\mathrm{top}$ computed for an infinite system with $L\to\infty$. Due to the finite value of $\beta$, regions with a finite topological gap of maximally $\Delta_\mathrm{top}\approx23$ $\mu$eV emerge even in the absence of Rashba SOC [Fig.~\ref{fig:PhaseDiagramInAs}(a)]. As $\alpha$ is increased, the extent of these regions and the maximal topological gap also increase: to $\Delta_\mathrm{top}\approx29$ $\mu$eV for $\alpha=4$ meVnm [Fig.~\ref{fig:PhaseDiagramInAs}(b)] and then to $\Delta_\mathrm{top}\approx38$ $\mu$eV for $\alpha=16$ meVnm [Fig.~\ref{fig:PhaseDiagramInAs}(c)]. These values of $\Delta_\mathrm{top}$ imply Majorana bound states that are sufficiently localized to be probed by tunneling spectroscopy in finite structures with $L$ of a few $\mu$m. Importantly, Fig.~\ref{fig:PhaseDiagramInAs}(b) again illustrates the disappearance of $\Delta_\mathrm{top}$ and thus Majorana bound states at $\phi=\pi$ for $\alpha=\beta$.

Simulations for InSb quantum wells give qualitatively similar results.\footnote{For InSb quantum wells, we use $m=0.0135m_0$, $\Delta=250$ $\mu$eV, $\mu_\mathrm{S}=1$ meV, $\mu_\mathrm{N}=0.7$ meV, $g=25$, $W=100$ nm, and $W_\mathrm{S}=450$ nm. Following Ref.~\onlinecite{Fabian2007:APS}, we take $\beta=32.4$ meVnm. The strength of Rashba SOC is then tuned from $\alpha=0$ to $\alpha=\beta$.} Here, the larger $g$-factor of InSb can lead to a topological transition already at smaller magnetic fields than in Fig.~\ref{fig:PhaseDiagramInAs}. Alternatively, this larger $g$-factor allows also for narrower junctions with an extended topological phase at accessible magnetic fields: For example, we find topological gaps of up to $\Delta_\mathrm{top}=90$ $\mu$eV well below $B=1$ T in InSb junctions with $W=100$ nm and $\Delta=250$ $\mu$eV (assuming $g=25$).

In addition to testing the disappearance and reappearance of Majorana bound states at $\phi=\pi$ by tuning to $\alpha=\beta$, other predictions could also be checked in InAs\footnote{In addition to the usual zinc-blende phase, the wurzite phase might also offer opportunities to engineer Majorana platforms based on InAs [P. E. Faria Junior, T. Campos, C. M. O. Bastos, M. Gmitra, J. Fabian, and G. M. Sipahi, Phys. Rev. B \textbf{93}, 235204 (2016); T. Campos, P. E. Faria Junior, M. Gmitra, G. M. Sipahi, and J. Fabian, Phys. Rev. B \textbf{97}, 245402 (2018)].} or InSb quantum wells. For example, the importance of the in-plane crystallographic axis along which the Josephson junction is set up could be verified by comparing the edge LDOS of the [100] and [110] setups if $\alpha$ is tuned to $\alpha=-\beta$: In [100] junctions, Majorana bound states can be expected for $\phi\neq\pi$, similar to the case $\alpha=\beta$ discussed above. In [110] junctions, on the other hand, Majorana bound states will be absent not only for $\phi=\pi$, but for any phase difference $\phi$. Another testable prediction is that the orientation of $\bm{B}$ has to be adjusted to the combination of $\alpha$ and $\beta$ in junctions with phase bias along the [100] direction: Rotating the in-plane field in the N region for a fixed combination of $\alpha$ and $\beta$ is expected to yield signatures of Majorana bound states only if $\bm{B}$ does not significantly deviate from $\bm{B}\perp\bm{n}_\mathrm{soc}$. As the orientation of $\bm{B}$ is rotated, Majorana bound states will disappear.

The above examples show that allowing for Dresselhaus SOC in phase-controlled Josephson junctions does not only extend topological superconductivity to a wide class of 2DEGs, but also adds another tunable knob. Hence, junctions with mixed Rashba and Dresselhaus SOC present a versatile platform for topological superconductivity with high tunability. Finally, the stability of this setup is an important point to consider. The small proximity-induced superconducting and topological gaps, $\Delta$ and $\Delta_\mathrm{top}$, require working at low temperatures.\cite{Ren2019:N,Fornieri2019:N} For example, resolving gaps of 40 $\mu$eV, that is, the maximal gap in Fig.~\ref{fig:PhaseDiagramInAs}, sets an upper boundary for the temperature of around 500 mK. To resolve also all the smaller gaps in Fig.~\ref{fig:PhaseDiagramInAs}, one should consequently work at temperatures of a few tens of mK. Another issue concerns the role of disorder: Although strong disorder eventually destroys the topological phase, it has recently been shown that weak disorder can actually be beneficial for the stability of quasi-1D topological superconductors, such as phase-controlled Josephson junctions.\cite{Haim2019:PRL} Therefore, we expect the system studied here to be stable against weak disorder and at low temperatures in the range of a few tens to a few hundreds of mK.

\section{Conclusions}\label{Sec:Conclusions}

We have studied Josephson junctions based on quantum wells with [001] growth direction and strong spin-orbit coupling subject to an in-plane magnetic field. For an arbitrary combination of Rashba and Dresselhaus spin-orbit coupling, a topological phase hosting Majorana bound states at the ends of the normal region can emerge for a wide range of parameters (chemical potential, superconducting phase difference $\phi$, strength of the magnetic field/Zeeman term $\bm{E}_\mathrm{Z}$). This topological phase forms if the magnetic field is oriented perpendicular to the spin-orbit field $\bm{n}_\mathrm{soc}$, which is defined by propagation parallel to the superconducting/normal interfaces. Hence, quasi-1D topological superconductivity based on phase-controlled Josephson junctions can appear in a wide class of two-dimensional electron gases grown along the [001] direction.

In $\bm{E}_\mathrm{Z}$-$\phi$ space, this topological phase emerges in extended regions inside diamonds centered around the Thouless energy and $\phi=\pi$ (or, more generally, in regions centered around odd multiples of the Thouless energy and of $\phi=\pi$, see App.~\ref{Sec:ScattComp}). The topological gap protecting the Majorana bound states is not only limited by the width of the normal region, but depends on the details of spin-orbit coupling, the Zeeman energies, and the superconducting phase difference. We have determined a topological phase diagram to find the parameter ranges most promising for the observation of well-localized Majorana bound states in narrow junctions: Typically, extended topological regions emerge in the limit of large spin-orbit coupling compared to the Zeeman energies, $\lambda_\mathrm{soc}k_\mathrm{F,N}\gg|\bm{E}_\mathrm{Z}|$. Surprisingly, for equal Rashba and Dresselhaus spin-orbit coupling, well-localized Majorana bound states can appear only for phase differences $\phi\neq\pi$ as the topological gap vanishes at $\phi=\pi$.

Based on our calculations, Dresselhaus spin-orbit coupling offers an additional knob to test Majorana bound states in phase-controlled Josephson junctions, either by tuning the ratio between Rashba and Dresselhaus spin-orbit coupling and/or by rotating the in-plane crystallographic axis along which the phase bias is applied. Finally, measurements of the local density of states at the edge of the normal region enable one to reconstruct (at least parts of) the topological phase diagram in narrow as well as in wide junctions. Future research directions could involve studying the Doppler shift which can arise due to magnetic-field-induced local gradients of the superconducting phase in the junctions discussed here.\cite{Rohlfing2009:PRB,Tkachov2015:PRB}

\acknowledgments
We thank Bertrand Halperin, Michael Kosowsky, Andrew Saydjari, Julian-Benedikt Mayer, and Florian Goth for valuable discussions. B.S. and E.M.H. acknowledge financial support by the German Research Foundation (DFG) through SFB 1170 ``ToCoTronics'' and through the W\"urzburg-Dresden Cluster of Excellence on Complexity and Topology in Quantum Matter -- \textit{ct.qmat} (EXC 2147, project-id 39085490) and by the ENB Graduate School on Topological Insulators. F.P. acknowledges financial support by the STC Center for Integrated Quantum Materials under NSF grant No. DMR-1231319. H.R. acknowledges funding provided by the Institute for Quantum Information and Matter, an NSF Physics Frontiers Center, NSF Grant No. PHY-1733907. A.Y. acknowledges funding by the NSF DMR-1708688, the STC Center for Integrated Quantum Materials under NSF grant No. DMR-1231319, the NSF GRFP under grant DGE1144152, and the US Army Research Office under grant W911NF-18-1-0316.

\quad

\quad

\appendix

\section{Scattering approach}\label{Sec:Scatt}
\subsection{Dispersion}\label{Sec:ScattDisp}
Complementary to our numerical finite-difference calculations (see below, Appendix~\ref{Sec:FD}), we also employ a scattering approach, where simple analytical expressions can be obtained in some limiting cases. One such case is the $\delta$-barrier model, which will be discussed here. As mentioned in Sec.~\ref{Sec:ky0}, the Hamiltonian at $k_y=0$ can be mapped to a Hamiltonian with $\beta=0$ and arbitrary Zeeman field. Hence, we briefly discuss a scattering approach for this case of $\alpha\neq0$, $\beta=0$, arbitrary $\bm{E}_\mathrm{Z}$, and a $\delta$-barrier. To start with, we for the moment keep a finite momentum $k_y$ and will only invoke $k_y=0$ later. Moreover, we adopt the slightly modified phase convention $\Phi(x)=\Theta(x)\phi$ to describe the superconducting phase difference $\phi$ between the two S regions.

Making use of translational invariance along the $y$-direction and choosing the ansatz $\Psi(x,y)=\e^{\i k_yy}\psi_{k_y}(x)/\sqrt{S}$, where $\psi_{k_y}(x)$ is a spinor in Nambu space and $S$ is the unit area, the Andreev bound states ($|E|<\Delta$) for Eqs.~(\ref{eq:BDGHam}) and~(\ref{eq:BDG}) can be described by the ansatz
\begin{widetext}
\begin{equation}\label{eq:SRansatz}
\psi_{k_y}(x)=\frac{1}{\sqrt{2}}\left\{\begin{array}{l}
\sum\limits_{\sigma=\pm}\left[a_{1,\sigma}\left(\begin{array}{l}1\\B_{e,\sigma,-}\\C_-\\C_-B_{e,\sigma,-}\end{array}\right)\e^{-\i q_{e,\sigma}x}+a_{2,\sigma}\left(\begin{array}{l}1\\B_{h,\sigma,+}\\C_+\\C_+B_{h,\sigma,+}\end{array}\right)\e^{\i q_{h,\sigma}x}\right],\; x<0,\\
\quad\\
\quad\\
\sum\limits_{\sigma=\pm}\left[b_{1,\sigma}\left(\begin{array}{l}1\\ B_{e,\sigma,+}\\\e^{-\i\phi}C_-\\\e^{-\i\phi}C_-B_{e,\sigma,+}\end{array}\right)\e^{\i q_{e,\sigma}x}+b_{2,\sigma}\left(\begin{array}{l}1\\ B_{h,\sigma,-}\\\e^{-\i\phi}C_+\\\e^{-\i\phi}C_+B_{h,\sigma,-}\end{array}\right)\e^{-\i q_{h,\sigma}x}\right],\; x>0.
\end{array}\right.
\end{equation}

Here, $C_\pm=(E\pm\i\sqrt{\Delta^2-E^2})/\Delta$ and the index $\sigma=\pm$ refers to the two chiralities of the Rashba system with
\begin{equation}\label{eq:SRansatzDef}
\begin{array}{l}
q_{e/h,\sigma}=\sqrt{\left[\frac{\sqrt{2m(\mu_\mathrm{S}\pm\i\sqrt{\Delta^2-E^2})}}{\hbar}-\sigma k_\mathrm{soc}\right]^2-k_y^2},\\
B_{e,\sigma,\pm}=\i\sigma\frac{\pm q_{e,\sigma}+\i k_y}{\sqrt{q_{e,\sigma}^2+k_y^2}},\quad B_{h,\sigma,\pm}=\i\sigma\frac{\pm q_{h,\sigma}+\i k_y}{\sqrt{q_{h,\sigma}^2+k_y^2}}
\end{array}
\end{equation}
with $k_\mathrm{soc}=m\alpha/\hbar^2=m\lambda_\mathrm{soc}/\hbar^2$. For a finite N region, the states inside the N region $-W/2<0<W/2$ would also have to be considered.

The ansatz~(\ref{eq:SRansatz}) is chosen such that $\lim\limits_{|x|\to\infty}|\psi_{k_y}(x)|^2=0$ and the 8 coefficients $a_{1,\sigma}$, $a_{2,\sigma}$, $b_{1,\sigma}$, $b_{2,\sigma}$ as well as the energy $E$ of the Andreev bound states can be determined from the boundary conditions at the $\delta$-barrier at $x=0$,
\begin{equation}\label{eq:SRBC}
\begin{array}{l}
\psi_{k_y}(0^+)=\psi_{k_y}(0^-),\\
\quad\\
\partial_x\psi_{k_y}(0^+)-\partial_x\psi_{k_y}(0^-)=\frac{2mW}{\hbar^2}\left(V_0-\bm{E}_\mathrm{Z}\cdot\bm{s}\,\tau_z\right)\psi_{k_y}(0).
\end{array}
\end{equation}
These boundary conditions then lead to a linear system of equations for the coefficients $a_{1,\sigma}$,...,$b_{2,\sigma}$, which we require to be nontrivial. This requirement, in turn, provides a condition from which the energy $E$ can be extracted. In principle, Eqs.~(\ref{eq:SRansatz})-(\ref{eq:SRBC}) have to be solved numerically, and only in certain limits of the $\delta$-barrier model are compact analytical solutions possible. These cases will be discussed in the following.

The problem is greatly simplified in the case of a $\delta$-barrier with only $E_{\mathrm{Z},y}\neq0$ and $k_y=0$. In this case, $\left[\hat{H}_\mathrm{BdG}(k_y=0),s_y\right]=0$. Then, instead of having to look at all the exponentially decaying states in the S regions [4 states in each S region with 2 states from each chirality $\sigma$; see Eq.~(\ref{eq:SRansatz})], we have two different solutions
\begin{equation}\label{eq:SRansatzEZy}
\psi_s(x)=\frac{1}{\sqrt{2}}\left\{\begin{array}{l}
\left[a_{e,s}\left(\begin{array}{l}\chi_s\\C_-\chi_s\end{array}\right)\e^{-\i\tilde{q}_ex}+a_{h,s}\left(\begin{array}{l}\chi_s\\C_+\chi_s\end{array}\right)\e^{\i\tilde{q}_hx}\right]\e^{-\i sk_\mathrm{soc}x},\; x<0,\\
\quad\\
\left[b_{e,s}\left(\begin{array}{l}\chi_s\\\e^{-\i\phi}C_-\chi_s\end{array}\right)\e^{\i\tilde{q}_ex}+b_{h,s}\left(\begin{array}{l}\chi_s\\\e^{-\i\phi}C_+\chi_s\end{array}\right)\e^{-\i\tilde{q}_hx}\right]\e^{-\i sk_\mathrm{soc}x},\; x>0.\\
\end{array}\right.
\end{equation}
\end{widetext}
Here, $s=\pm$ refers to the two eigenvalues of the spin Pauli matrix $s_y$ with the corresponding spinors $\chi_s$ and
\begin{equation}
\tilde{q}_{e/h}=\frac{\sqrt{2m(\mu_\mathrm{S}\pm\i\sqrt{\Delta^2-E^2})}}{\hbar}.
\end{equation}
Invoking the boundary conditions~(\ref{eq:SRBC}) and requiring a nontrivial solution for the coefficients $a_{e/h,s}$ and $b_{e/h,s}$ yields a transcendental equation for the energy $E$. While this transcendental equation cannot be solved analytically without further approximations, we can determine whether zero-energy solutions are possible and at which phase differences they appear.

One can find zero-energy solutions at phase differences $\phi$, which for both $s=\pm$ satisfy
\begin{equation}\label{eq:SRzee}
\begin{array}{ll}
\cos\phi=&\frac{k_\mathrm{F,S}^2-3\sqrt{k_\mathrm{F,S}^4+\kappa^4}}{k_\mathrm{F,S}^2+\sqrt{k_\mathrm{F,S}^4+\kappa^4}}\\
&\quad\quad+\frac{k_\mathrm{F,S}^2Z_y^2-k_\mathrm{F,S}^2Z_0^2-4k_\mathrm{F,S}Z_0\mathrm{Im}\left(\sqrt{k_\mathrm{F,S}^2+\i\kappa^2}\right)}{k_\mathrm{F,S}^2+\sqrt{k_\mathrm{F,S}^4+\kappa^4}},
\end{array}
\end{equation}
where we have introduced $k_\mathrm{F,S}^2=2m\mu_\mathrm{S}/\hbar^2$, $\kappa^2=2m\Delta/\hbar^2$, $Z_y=2mE_{\mathrm{Z},y}W/(\hbar^2k_\mathrm{F,S})=2E_{\mathrm{Z},y}W/(\hbar v_F)$, and $Z_0=2mV_0W/(\hbar^2k_\mathrm{F,S})$. Equation~(\ref{eq:SRzee}) makes it clear that even for $V_0=E_{\mathrm{Z},y}=0$ (that is, $Z_y=Z_0=0$) this equation cannot be solved since
\begin{equation}
\frac{k_\mathrm{F,S}^2-3\sqrt{k_\mathrm{F,S}^4+\kappa^4}}{k_\mathrm{F,S}^2+\sqrt{k_\mathrm{F,S}^4+\kappa^4}}\leq-1.
\end{equation}
Only in the limit of $\kappa\to0$, one can obtain $\phi=n\pi$ with $n\in\mathbb{Z}$. Hence, because of scattering between the two chiralities (due to the finite $\Delta$) a gap is opened at $\phi=n\pi$. A finite $V_0$ does not change this picture. However, a finite $E_{\mathrm{Z},y}$ can enable a solution of Eq.~(\ref{eq:SRzee}).

This is most clearly seen in the Andreev approximation, $\Delta\ll\mu_\mathrm{S}$, when Eq.~(\ref{eq:SRzee}) reduces to
\begin{equation}\label{eq:SRzeeA}
\cos\phi=-1+\frac{Z_y^2-Z_0^2}{2},
\end{equation}
compare also Eq.~(\ref{eq:DAbc}). This equation explicitly shows the competition between $Z_y$ and $Z_0$, that is, between $E_{\mathrm{Z},y}$ and $V_0$. For $Z_0\neq0$ and $Z_y=0$, Eq.~(\ref{eq:SRzeeA}) cannot be solved. For finite $Z_y$, on the other hand, a solution is possible as long as $Z_y^2>Z_0^2$ and $Z_y^2-Z_0^2<4$. This last requirement makes it clear that, although a finite $Z_y$ restores a zero-energy solution, $Z_y$ cannot become too large. Otherwise, no zero-energy solution is possible. This behavior of a $\delta$-junction can also be found in narrow finite S/N/S junctions.

Whereas closed analytical expressions for the dispersion cannot be obtained for general $\mu_\mathrm{S}$ and $\Delta$, we can obtain the dispersion in the Andreev approximation. In this case, each spin $s=\pm$ yields two solutions given by
\begin{equation}\label{eq:SRAEZy}
E_{s,\pm}(k_y=0,\phi)=\frac{\sgn\left[f_\pm^s(Z_0,Z_y,\phi)\right]\Delta}{\sqrt{1+\left[f_\pm^s(Z_0,Z_y,\phi)\right]^2}},
\end{equation}
where
\begin{equation}\label{eq:SRAEZyDef}
\begin{array}{l}
f_\pm^s(Z_0,Z_y,\phi)=\\
\quad\quad\quad\frac{2sZ_y\pm\sqrt{2\left(1+Z_0^2+Z_y^2+(Z_y^2-Z_0^2)\cos\phi-\cos2\phi\right)}}{2+Z_0^2-Z_y^2+2\cos\phi}.
\end{array}
\end{equation}

It is not possible to obtain compact analytical expressions for a general $\bm{E}_\mathrm{Z}$ in a $\delta$-barrier model. However, using the boundary conditions~(\ref{eq:SRBC}) and the Andreev approximation, $\Delta\ll\mu_\mathrm{S}$, we can find the dispersion for $k_y=0$ as
\begin{equation}\label{eq:SRAEZ}
\left|E(k_y=0,\phi)\right|=\frac{\Delta}{\sqrt{1+\left[f_\pm^+(Z_0,Z_\mathrm{Z},\phi)\right]^2}},
\end{equation}
where $Z_\mathrm{Z}^2=Z_x^2+Z_y^2+Z_z^2$. Note that for $Z_y\neq0$ and $Z_x=Z_z=0$, the four solutions given by Eq.~(\ref{eq:SRAEZ}) are equivalent to the solutions given by Eq.~(\ref{eq:SRAEZy}) if both spin eigenvalues of $s_y$, $s=\pm$, are considered. Defining $f_\pm(Z_0,Z_\mathrm{Z},\phi)\equiv f_\pm^+(Z_0,Z_\mathrm{Z},\phi)$, Eq.~(\ref{eq:SRAEZ}) is Eq.~(\ref{eq:DAdisp}) found in the main text.

\begin{figure}[t]
\centering
\includegraphics*[width=8.5cm]{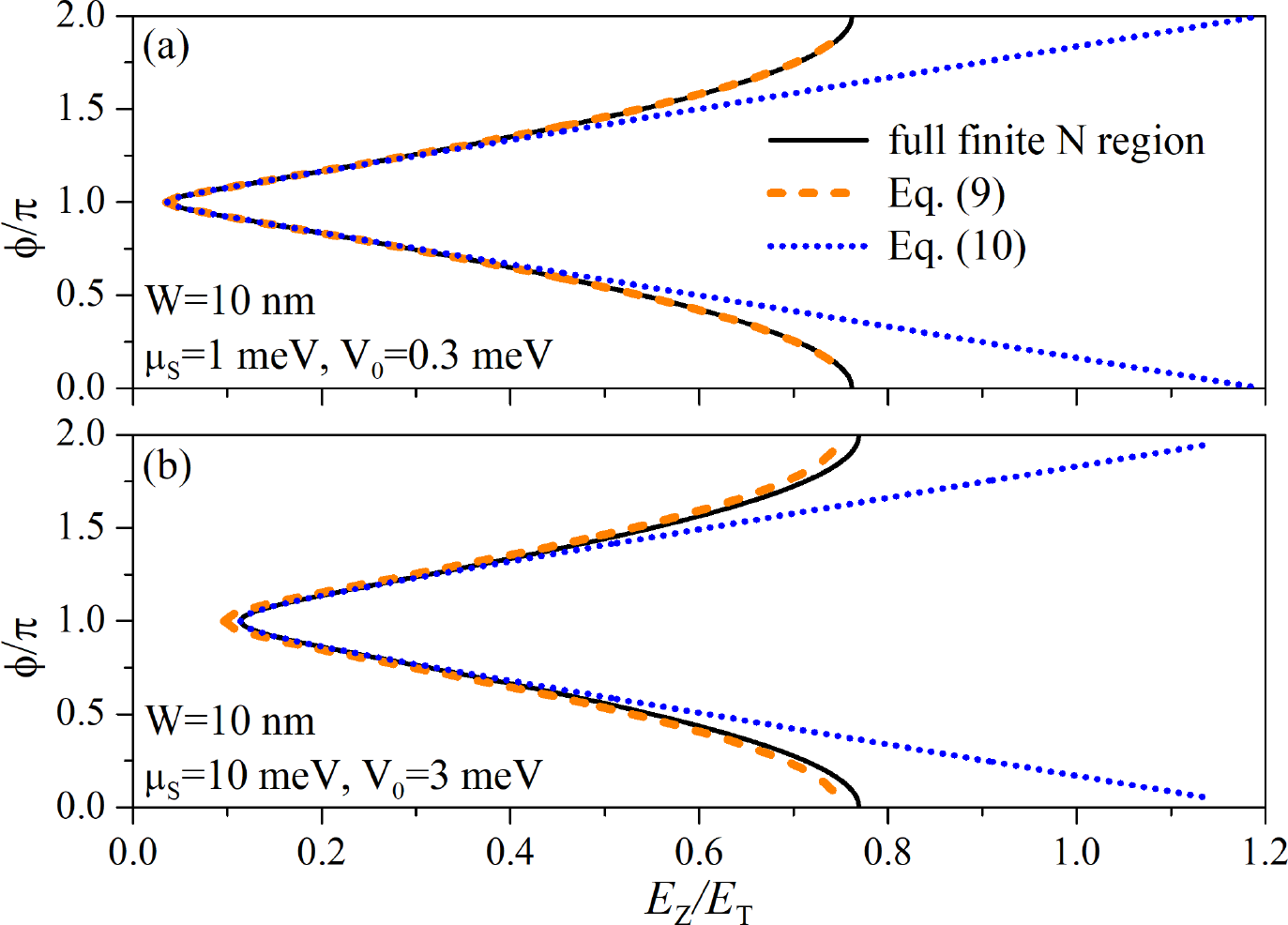}
\caption{(Color online) Phase boundaries of a ultra-narrow junction of width $W=10$ nm obtained from the scattering approach for a finite barrier and a $\delta$-barrier model. In panel~(a), $\mu_\mathrm{S}=1$ meV and $V_0=0.3$ meV, while $\mu_\mathrm{S}=10$ meV and $V_0=3$ meV in panel~(b). Here, the S regions are assumed to be semi-infinite in the $x$ direction. The other parameters are $m=0.038m_0$, $\Delta=250$ $\mu$eV, and $\bm{E}_\mathrm{Z}\perp\bm{n}_\mathrm{soc}$.}\label{fig:CompDeltaFinite10nm}
\end{figure}

\subsection{Phase boundaries and comparison between the $\delta$-barrier and a junction with finite normal region}\label{Sec:ScattComp}

From Eq.~(\ref{eq:DAdisp}), we can derive approximate expressions for the phase boundaries in the $\delta$-barrier model, Eqs.~(\ref{eq:DAbc}) or~(\ref{eq:DAZ}). As mentioned in the main text, a major drawback of the $\delta$-barrier model is that it describes only one phase transition, but not the multiple phase transitions between trivial and topological regimes encountered in the finite-barrier model. Still, one advantage of the $\delta$-barrier model is that it yields relatively simple expressions that capture many qualitative features found in junctions with finite normal regions.

Particularly for very narrow normal regions, the $\delta$-barrier model provides a good approximation of the finite-barrier model if $|\bm{E}_\mathrm{Z}|$ is not much smaller than $\mu_\mathrm{S}$. This is illustrated by Fig.~\ref{fig:CompDeltaFinite10nm}, where the phase boundaries obtained from a finite-barrier model and the $\delta$-barrier model, Eqs.~(\ref{eq:DAbc}) or~(\ref{eq:DAZ}), are shown for a very narrow junction with $W=10$ nm, $m=0.038m_0$, $\Delta=250$ $\mu$eV, and $\bm{E}_\mathrm{Z}\perp\bm{n}_\mathrm{soc}$. Here, the phase boundaries for the finite-barrier model, given by $E(k_y=0,\phi=\phi_c)=0$, have also been computed from a scattering approach with semi-infinite S regions in the $x$ direction. In this approach, the wave functions in the S regions are given by Eqs.~(\ref{eq:SRansatz}) like in the $\delta$-barrier model. Now however, the wave function in the finite N region is also taken into account, and we use the continuity of the wave function at the S/N interfaces instead of the matching conditions~(\ref{eq:SRBC}).

For small chemical potentials $\mu_\mathrm{S}$, there is almost perfect agreement between the full finite-barrier solution and the $\delta$-barrier solution~(\ref{eq:DAbc}) [Fig.~\ref{fig:CompDeltaFinite10nm}(a)]. Even the approximated $\delta$-barrier solution~(\ref{eq:DAZ}) describes the phase transition in this ultra-narrow junction very well around $\phi=\pi$. There are slight deviations between the solutions of a finite and a $\delta$-barrier if $\mu_\mathrm{S}$ is increased [Fig.~\ref{fig:CompDeltaFinite10nm}(b)]. Nevertheless, Eq.~(\ref{eq:DAbc}) remains a good approximation also for the finite barrier.

\begin{figure}[t]
\centering
\includegraphics*[width=8.5cm]{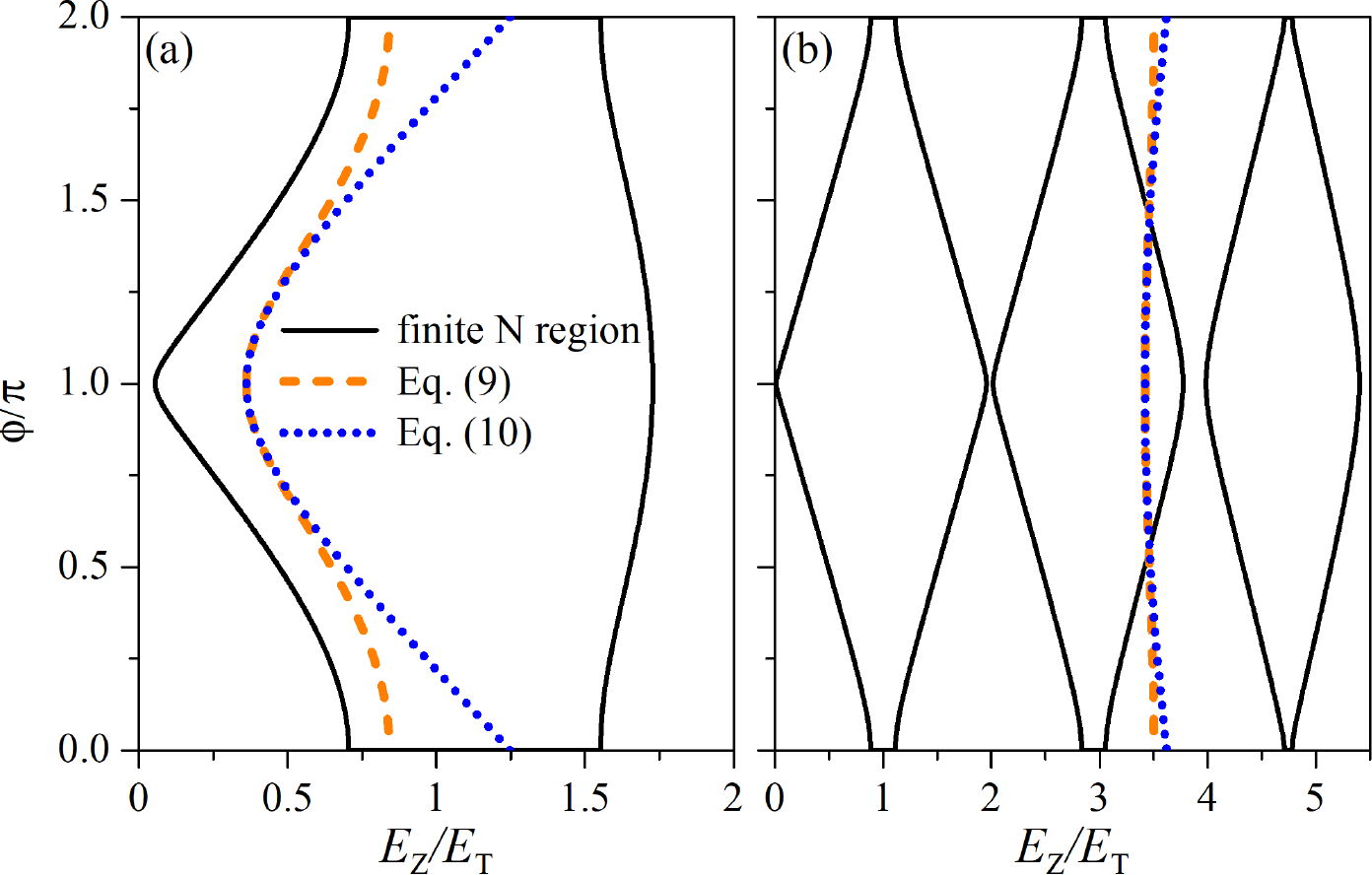}
\caption{(Color online) Phase boundaries of (a) a junction of width $W=100$ nm and (b) one of width $W=300$ nm obtained from the scattering approach for a finite barrier and a $\delta$-barrier model. In panel~(a), $\mu_\mathrm{S}=1$ meV and $V_0=0.3$ meV, while $\mu_\mathrm{S}=10$ meV and $V_0=3$ meV in panel~(b). All other parameters are similar to Fig.~\ref{fig:CompDeltaFinite10nm}.}\label{fig:CompDeltaFinite100vs300nm}
\end{figure}

If we go beyond the limit of ultra-narrow junctions to widths of a few hundred nanometers, the $\delta$-barrier solution no longer provides a very good description for the first phase transition. In Fig.~\ref{fig:CompDeltaFinite100vs300nm}, we again show a comparison between the phase boundaries obtained from the finite- and $\delta$-barrier models for junctions with $W=100$ nm and $W=300$ nm.

In fact, the parameters with $W=100$ nm [Fig.~\ref{fig:CompDeltaFinite100vs300nm}(a)] are the same as in Fig.~\ref{fig:PhaseDiagram} with the exception that now the S regions are semi-infinite instead of having a finite width $W_\mathrm{S}=450$ nm. Since $W_\mathrm{S}$ and $\Delta$ have been chosen large enough such that there is no significant normal reflection due to the finite width of the superconductors, the phase boundaries of the topological regions are almost identical in Fig.~\ref{fig:PhaseDiagram} (obtained from a finite-difference calculation) and Fig.~\ref{fig:CompDeltaFinite100vs300nm}(a) for the finite-barrier model (obtained from a scattering/wave-matching approach). The $\delta$-barrier model, on the other hand, gives a topological phase transition that occurs at higher magnetic fields than in the more realistic model with a finite N region. Furthermore, the transition to a trivial regime above $|\bm{E}_Z|=1.5$ meV is also not captured as well as the reentrance of a topological phase at even higher Zeeman energies (not shown here).

As $W$ is further increased, the deviation between the finite and $\delta$-barrier models becomes even more pronounced as illustrated in Fig.~\ref{fig:CompDeltaFinite100vs300nm}(b) for $W=300$ nm. Here, the single phase boundary occurs at values of the Zeeman energy where the finite system has already undergone several phase transitions. For comparable values of $|\bm{E}_\mathrm{Z}|$, a Josephson junction with finite $W$ is already in the topological phase described by a second diamond centered around $|\bm{E}_Z|=3E_T$. These additional phase transitions between trivial and topological regimes cannot be captured by the $\delta$-barrier model.

\section{Finite-difference method}\label{Sec:FD}

We employ a finite-difference method to solve the BdG Eq.~(\ref{eq:BDG}) with the Hamiltonians~(\ref{eq:BDGHam}) or~(\ref{eq:BDGHam_o110}) for a Josephson junction with finite S regions of widths $W_\mathrm{S}$ and a finite N region of width $W$ numerically. This finite-difference method is used for systems that are either infinite or confined in the $y$ direction.

If we first turn to the case of an infinite system in $y$ direction, we can introduce the good quantum number $k_y$ with the wave function $\Psi(x,y)=\e^{\i k_yy}\psi_{k_y}(x)/\sqrt{S}$. Hence, we only need to employ a 1D finite-difference scheme for the spinor $\psi_{k_y}(x)$, for which we invoke hard-wall boundary conditions at $x=\pm(W/2+W_\mathrm{S})$. We discretize the resulting differential equation for $\psi_{k_y}(x)$ and introduce $x\to x_l=a_xl-(W/2+W_\mathrm{S})$ with $l=0,1,...,N_x$, the number of discrete steps $N_x$ and the discrete step size $a_x=(W+2W_\mathrm{S})/N_x$. Using
\begin{equation}\label{eq:FDx}
\begin{array}{l}
\frac{\d\psi_{k_y}(x)}{\d x}\to\frac{\psi_{k_y}(x_l+a_x)-\psi_{k_y}(x_l-a_x)}{2a_x},\\
\frac{\d^2\psi_{k_y}(x)}{\d x^2}\to\frac{\psi_{k_y}(x_l+a_x)-2\psi_{k_y}(x_l)+\psi_{k_y}(x_l-a_x)}{a_x^2},
\end{array}
\end{equation}
the BdG $\hat{H}_\mathrm{BdG}(k_y)\psi_{k_y}(x)=E\psi_{k_y}(x)$ becomes
\begin{equation}\label{eq:BdGFDx}
\sum\limits_{l'=0}^{N_x}\mathcal{H}_{l,l'}(k_y)\psi_{k_y}(x_{l'})=E\psi_{k_y}(x_l)
\end{equation}
with
\begin{widetext}
\begin{equation}\label{eq:HxFD}
\begin{array}{ll}
\mathcal{H}_{l,l'}(k_y)=&\Bigg\{\left[\frac{\hbar^2}{ma_x^2}+\frac{\hbar^2k_y^2}{2m}-\alpha k_ys_x-\beta k_ys_y+\frac{m\left(\alpha^2+\beta^2\right)}{2\hbar^2}-\mu_\mathrm{S}\right]\tau_z+\left(V_0\tau_z-\bm{E}_\mathrm{Z}\cdot\bm{s}\right)h(x_l)\\
&\quad+\Delta(x_l)\left[\tau_x\cos\Phi(x_l)-\tau_y\sin\Phi(x_l)\right]\Bigg\}\delta_{l,l'}-\left[\frac{\hbar^2}{2ma_x^2}+\left(\alpha s_y+\beta s_x\right)\frac{\i\left(l'-l\right)}{2a_x}\right]\tau_z\left(\delta_{l,l'+1}+\delta_{l,l'-1}\right)
\end{array}
\end{equation}
for the Hamiltonian~(\ref{eq:BDGHam}) and
\begin{equation}\label{eq:HxFD_o110}
\begin{array}{ll}
\mathcal{H}_{l,l'}(k_y)=&\Bigg\{\left[\frac{\hbar^2}{ma_x^2}+\frac{\hbar^2k_y^2}{2m}-\left(\alpha+\beta\right)k_ys_x+\frac{m\left(\alpha^2+\beta^2\right)}{2\hbar^2}-\mu_\mathrm{S}\right]\tau_z+\left(V_0\tau_z-\bm{E}_\mathrm{Z}\cdot\bm{s}\right)h(x_l)\\
&\quad+\Delta(x_l)\left[\tau_x\cos\Phi(x_l)-\tau_y\sin\Phi(x_l)\right]\Bigg\}\delta_{l,l'}-\left[\frac{\hbar^2}{2ma_x^2}+\left(\alpha-\beta\right)\frac{\i\left(l'-l\right)}{2a_x}s_y\right]\tau_z\left(\delta_{l,l'+1}+\delta_{l,l'-1}\right)
\end{array}
\end{equation}
for the Hamiltonian~(\ref{eq:BDGHam_o110}). Here, $h(x_l)=\Theta(W/2-|x_l|)$, $\Delta(x_l)=\Delta\Theta(|x_l|-W/2)$, and $\Phi(x_l)=\phi\Theta(x_l-W/2)$.
\end{widetext}

For the case of a finite system in both the $x$ and $y$ direction, we now employ hard-wall boundary conditions at both $x=\pm(W/2+W_\mathrm{S})$ and $y=\pm L/2$. In this case, we can no longer introduce a good quantum number $k_y$. Instead, we now discretize $\Psi(x,y)$ in the $x$ as well as in the $y$ direction according to $x\to x_l=a_xl-(W/2+W_\mathrm{S})$ with $l=0,1,...,N_x$ and $y\to y_m=a_ym-L/2$ with $m=0,1,...,N_y$. The discrete step sizes are $a_x=(W+2W_\mathrm{S})/N_x$ and $a_y=L/N_y$. Rewriting the derivatives with respect to $x$ and $y$ similarly to Eqs.~(\ref{eq:FDx}), the BdG $\hat{H}_\mathrm{BdG}\Psi(x,y)=E\Psi(x,y)$ becomes
\begin{equation}\label{eq:BdGFDxy}
\sum\limits_{l'=0}^{N_x}\sum\limits_{m'=0}^{N_y}\mathcal{H}_{l,m;l',m'}\Psi(x_{l'},y_{m'})=E\Psi(x_l,y_m)
\end{equation}
with the Hamiltonians
\begin{widetext}
\begin{equation}\label{eq:HxyFD}
\begin{array}{l}
\mathcal{H}_{l,m;l',m'}=\\
\quad\Bigg\{\left[\frac{\hbar^2}{ma_x^2}+\frac{\hbar^2}{ma_y^2}+\frac{m\left(\alpha^2+\beta^2\right)}{2\hbar^2}-\mu_\mathrm{S}\right]\tau_z+\left(V_0\tau_z-\bm{E}_\mathrm{Z}\cdot\bm{s}\right)h(x_l)+\Delta(x_l)\left[\tau_x\cos\Phi(x_l)-\tau_y\sin\Phi(x_l)\right]\Bigg\}\delta_{l,l'}\delta_{m,m'}\\
\quad-\left[\frac{\hbar^2}{2ma_x^2}+\left(\alpha s_y+\beta s_x\right)\frac{\i\left(l'-l\right)}{2a_x}\right]\tau_z\left(\delta_{l,l'+1}+\delta_{l,l'-1}\right)\delta_{m,m'}-\left[\frac{\hbar^2}{2ma_y^2}-\left(\alpha s_x+\beta s_y\right)\frac{\i\left(m'-m\right)}{2a_y}\right]\tau_z\delta_{l,l'}\left(\delta_{m,m'+1}+\delta_{m,m'-1}\right)
\end{array}
\end{equation}
for Eq.~(\ref{eq:BDGHam}) and
\begin{equation}\label{eq:HxyFD_o110}
\begin{array}{l}
\mathcal{H}_{l,m;l',m'}=\\
\quad\Bigg\{\left[\frac{\hbar^2}{ma_x^2}+\frac{\hbar^2}{ma_y^2}+\frac{m\left(\alpha^2+\beta^2\right)}{2\hbar^2}-\mu_\mathrm{S}\right]\tau_z+\left(V_0\tau_z-\bm{E}_\mathrm{Z}\cdot\bm{s}\right)h(x_l)+\Delta(x_l)\left[\tau_x\cos\Phi(x_l)-\tau_y\sin\Phi(x_l)\right]\Bigg\}\delta_{l,l'}\delta_{m,m'}\\
\quad-\left[\frac{\hbar^2}{2ma_x^2}+\left(\alpha-\beta\right)\frac{\i\left(l'-l\right)}{2a_x}s_y\right]\tau_z\left(\delta_{l,l'+1}+\delta_{l,l'-1}\right)\delta_{m,m'}-\left[\frac{\hbar^2}{2ma_y^2}-\left(\alpha+\beta\right)\frac{\i\left(m'-m\right)}{2a_y}s_x\right]\tau_z\delta_{l,l'}\left(\delta_{m,m'+1}+\delta_{m,m'-1}\right)
\end{array}
\end{equation}
for Eq.~(\ref{eq:BDGHam_o110}). Again, $h(x_l)=\Theta(W/2-|x_l|)$, $\Delta(x_l)=\Delta\Theta(|x_l|-W/2)$, and $\Phi(x_l)=\phi\Theta(x_l-W/2)$.
\end{widetext}

In our calculations, we have chosen the step sizes $a_x=a_y=20$ nm. For the structures that are infinite in the $y$ direction and have the width $W_{tot}=1\mu$m in the $x$ direction, this corresponds to a 1D grid with 50 lattice points. For the full 2D structures with the width $W_{tot}=1\mu$m in the $x$ direction and the length $L=2\mu$m in the $y$ direction, these step sizes correspond to a 2D grid with $50\times100$ lattice points.

\bibliography{BibTopInsAndTopSup}

\end{document}